\documentclass[preprint2,trackchanges]{aastex63}

\accepted{June 14, 2021}
\submitjournal{ApJ}

\shorttitle{XMM175UND-Field observed by XMM--Newton}
\shortauthors{El\'ias-Ch\'avez et al.}
\graphicspath{{./}{figures/}}
\usepackage{multirow}
\usepackage{graphicx}

\graphicspath{{./}{figures/}}
\usepackage{graphicx}	% Including figure files
\usepackage[utf8]{inputenc}
\usepackage{wrapfig}
\usepackage{tablefootnote}
\usepackage{amsmath}	% Advanced maths commands
\usepackage{amssymb}	% Extra maths symbols

\defcitealias{Traulsen2020}{DR9s}
\defcitealias{Webb2020}{DR9}
\defcitealias{Chen2018}{Chen}

\begin{document}

\title{X-ray sources in the 1.75 Ms Ultra Narrow Deep Field observed by XMM-Newton }

\correspondingauthor{Mauricio El\'ias Ch\'avez}
\email{melias@inaoep.mx}

\author[0000-0002-0115-8374]{M. El\'ias-Ch\'avez}
\affiliation{Instituto Nacional de Astrof\'isica, \'Optica y Electr\'onica, Luis E. Erro 1, Tonantzintla, Puebla, C.P. 72840, Mexico}

\author[0000-0001-8825-3624]{A. L. Longinotti}
\affiliation{Instituto de Astronom\'ia, Universidad Nacional Aut\'onoma de M\'exico,  Circuito Exterior, Ciudad Universitaria, Ciudad de M\'exico 04510, M\'exico}
\affiliation{CONACyT}

\author[0000-0001-6291-5239]{Y. Krongold}
\affiliation{Instituto de Astronom\'ia, Universidad Nacional Aut\'onoma de M\'exico,  Circuito Exterior, Ciudad Universitaria, Ciudad de M\'exico 04510, M\'exico}

\author[0000-0002-8853-9611]{C. Vignali}
\affiliation{Dipartimento di Fisica e Astronomia, Universit\`a di Bologna, viale Berti Pichat 6/2, 40127 Bologna, Italy}
\affiliation{INAF - Osservatorio di Astrofisica e Scienza dello Spazio di Bologna, Via Gobetti 93/3, I-40129 Bologna, Italy}

\author[0000-0002-6896-1364]{F. Nicastro}
\affiliation{Observatorio Astronomico di Roma-INAF, Via di Frascati 33, 1-00040 Monte Porzio Catone, RM, Italy}

\author[0000-0003-1327-0838]{D. Rosa-Gonz\'alez}
\affiliation{Instituto Nacional de Astrof\'isica, \'Optica y Electr\'onica, Luis E. Erro 1, Tonantzintla, Puebla, C.P. 72840, Mexico}

\author[0000-0002-4677-0516]{Y. D. Mayya}
\affiliation{Instituto Nacional de Astrof\'isica, \'Optica y Electr\'onica, Luis E. Erro 1, Tonantzintla, Puebla, C.P. 72840, Mexico}

\author[0000-0002-4822-3559]{S. Mathur}
\affiliation{Department of Astronomy, The Ohio State University, 140 West 18th Avenue, Columbus, OH 43210, USA}
\affiliation{Center for Cosmology and Astroparticle Physics, 191 West Woodruff Avenue, Columbus, OH 43210, USA}

%% Note that the \and command from previous versions of AASTeX is now
%% depreciated in this version as it is no longer necessary. AASTeX 
%% automatically takes care of all commas and "and"s between authors names.

%% AASTeX 6.3 has the new \collaboration and \nocollaboration commands to
%% provide the collaboration status of a group of authors. These commands 
%% can be used either before or after the list of corresponding authors. The
%% argument for \collaboration is the collaboration identifier. Authors are
%% encouraged to surround collaboration identifiers with ()s. The 
%% \nocollaboration command takes no argument and exists to indicate that
%% the nearby authors are not part of surrounding collaborations.

%% Mark off the abstract in the ``abstract'' environment. 
\begin{abstract}

In this work we present the results of the survey carried out on one of the deepest X-ray fields observed by the XMM-Newton satellite. The 1.75 Ms Ultra Narrow Deep Field (XMM175UNDF) survey is made by 13 observations taken over 2 years with a total exposure time of 1.75 Ms  (1.372  Ms after flare-filtered) in a field of $30\arcmin \times 30\arcmin $ centered around the blazar 1ES 1553+113. We stacked the 13 observations reaching flux limits of $4.03 \times 10^{-16} $, $1.3 \times 10^{-15}$, and $9.8 \times 10^{-16}\, erg\, s^{-1}\, cm^{-2}$  in the soft $(0.2 - 2\, \mathrm{keV})$, hard  $(2 - 12\, \mathrm{keV})$, and full $(0.2 - 12\, \mathrm{keV})$ bands, respectively. Using a conservative threshold of Maximum Likelihood significance of $ML \geq 6$, corresponding to $3\sigma$, we detected 301 point-sources for which we derived positions, fluxes in different bands, and hardness ratios. Thanks to an optical follow-up carried out using the 10.4m the \textit{Gran Telescopio Canarias} (GTC) on the same field in the $u'g'r'i'z'$ bands, combined with WISE/2MASS IR data; we identified 244 optical/IR counterpart candidates for our X-ray sources and estimated their X-ray luminosities, redshift distribution, X-ray/optical $\--$ X-ray/IR  flux ratios, and absolute magnitudes. Finally, we divided this subsample in 40 non-active sources and 204 AGNs, of which 139 are classified as Seyfert galaxies and 41 as Quasars.

\end{abstract}

%% Keywords should appear after the \end{abstract} command. 
%% See the online documentation for the full list of available subject
%% keywords and the rules for their use.

%% From the front matter, we move on to the body of the paper.
%% Sections are demarcated by \section and \subsection, respectively.
%% Observe the use of the LaTeX \label
%% command after the \subsection to give a symbolic KEY to the
%% subsection for cross-referencing in a \ref command.
%% You can use LaTeX's \ref and \label commands to keep track of
%% cross-references to sections, equations, tables, and figures.
%% That way, if you change the order of any elements, LaTeX will
%% automatically renumber them.
%%
%% We recommend that authors also use the natbib \citep
%% and \citet commands to identify citations.  The citations are
%% tied to the reference list via symbolic KEYs. The KEY corresponds
%% to the KEY in the \bibitem in the reference list below. 

\section{Introduction} \label{sec:intro}

One of the biggest problems in cosmology is the understanding of the connection between Super Massive Black Holes (SMBHs) and Galaxy formation \citep{Merritt2000,DiMatteo2005,Done2010}. In order to uncover this co-evolution it is necessary to  detect and characterize large samples of Active Galactic Nuclei (AGNs) and their hosts, by using multi-wavelength analysis through larger and deeper surveys in bands as optical, infrared, radio,   and X-ray  \citep{Scoville2007,Kellermann2008,Rosen2016A,Brandt2017}. AGNs are galaxies that host an accreting SMBH in their nuclear region, which emits a large amount of X-ray photons via accretion processes \citep{Haardt1991,George1991,Matt1997}. 
Contrary to optical and infrared surveys, which may suffer incompleteness and/or misidentification problems \citep[e.g.][]{Scoville2007}, X-ray surveys provide a very powerful tool to blindly search for AGNs \citep{Brandt2015}. Additionally,  1)  X-ray emission can penetrate through high column densities of material  $(N_{H}  = 10^{21} - 10^{24.5}\,  \mathrm{cm^{-2}})$   allowing for the detection of moderately obscured AGNs \citep{Ghisellini1994,Ghosh2008,Hickox2018}; 2) X-ray emission of AGNs suffers low dilution by their host galaxy as opposed to radiation in the  optical band; 3) X-ray spectra of AGNs can be used as a diagnostic tool to infer luminosity, obscuration level, nuclear geometry, disk/corona conditions, and Eddington ratio $(L_{Bol}/L_{Edd})$ \citep{Brandt2017}. Therefore,  X-ray surveys allow us to identify  large samples of obscured $(Log\, N_{H} > 21.5\, \mathrm{cm^{-2}} )$ and unobscured AGNs, making possible to study their contribution to the Cosmic X-ray Background (XRB), associated to the integrated  X-ray emission from extragalactic faint point sources \citep{Gilli2007}.

In the last two decades X-ray missions such as XMM-Newton and Chandra performed shallow X-ray surveys over wide fields and  deep surveys in narrow areas \citep[for a detailed summary see][]{Brandt2015}. The strategy of surveying large areas is optimal to look into large volumes of the universe, increasing the probability to find high-luminous QSOs and atypical sources that could be missed by small coverage surveys \citep{Evans2010,Warwick2012,Rosen2016A}. In contrast,  deep X-ray surveys in narrow field areas are an effective method  to identify moderately luminous AGNs and faint high-redshift sources \citep{Brusa2007,Puccetti2009,Marchesi2016,Vito2016}.

The X-ray observations analyzed here were originally dedicated to study the Warm Hot Intergalactic Medium (WHIM) with the goal of observing highly ionized intervening absorbers via detection of OVII features in the spectrum of the blazar IES 1553+113 \citep{Nicastro2018,Das2019}. 

This project gathered in 2 years a total of 13 observations targeting the blazar and the  $30\arcmin \times 30\arcmin$ area around it, generating a total exposure time of 1.75 Ms.  As a by-product, this program created the 1.75 Ms Ultra Narrow Deep Field (XMM175UNDF), one of the narrowest and deepest surveys ever performed with
XMM-Newton in the band $0.2 - 12\, \mathrm{keV}$, particularly well-suited to survey the AGN content of the field. 

Furthermore, to search for optical counterparts and provide solid photometric and spectroscopic identifications, we performed an optical campaign of this field with the OSIRIS camera mounted at the 10m \textit{Gran Telescopio Canarias} (GTC). Finally, we cross-correlated our X-ray/optical catalog with available infrared (IR) coverage by WISE/2MASS from \citet{Cutri2013}. 

In this paper we present a catalog of 301 X-ray point-sources\footnote{The present XMM-Newton catalog with its optical (GTC) and Infrared (WISE/2MASS) counterpart associations, is publicly available for further analysis  in ASCII format along with this paper}, consistent  with the results obtained for this field by the XMM-Newton Survey Science Centre and recently reported by \citet{Webb2020, Traulsen2020}. 
 
This paper is organized as follows: in Section 2 we present the XMM-Newton observations and procedures for data reduction. We describe the method used to identify the X-ray sources and the details of the production of the X-ray point-source catalog and its statistical reliability. In Section 3 we identify the optical/IR counterparts  by cross-matching the X-ray catalogue with the optical/IR catalogues, using the Likelihood Ratio (LR) technique and we explain our photo-zs determination procedure. In Section 4 we describe the general properties of the X-ray catalog and the $Log\, N - Log\ S$ data analysis. In Section 5 we present the results of our multiwavelength analysis (e.g. luminosity distribution, AGN identifications). In Section 6 and 7 we discuss and summarize the most important results of the paper. Throughout this work, we adopted the cosmological parameters $H_0 = 70\, \mathrm{km\, s^{-1}\,  Mpc^{-1}} ,\Omega_m=0.3$ and $\Omega_\Lambda=0.7$.

\begin{table*}
\caption{ Resume of XMM-Newton observations  around the blazar 1ES 1553+113. The distance in the last column is measured in arcminute from the center of each observation to the blazar.}
\label{tabla1}
\centering
%\resizebox{\linewidth}{!}{%
\begin{tabular}{ccccccc} 
\hline \hline
Obs.ID & Date & Nominal Exp & Exp clean PN & Exp clean MOS1 & Exp clean MOS2 & Distance \\
 &  & $(ks)$ & $(ks)$ & $(ks)$ & $(ks)$~ & $(arcmin)$ \\ 
\hline
761100101 & 2015 Jul 29 & 138.4 & 126.3 & 128.5 & 133.6 & 0.5 \\
761100201 & 2015 Aug 2 & 138.9 & 122.1 & 130.4 & 128.6 & 0.24 \\
761100301 & 2015 Aug 4 & 138.9 & 133.4 & 131.5 & 135.5 & 0.25 \\
761100401 & 2015 Aug 8 & 138.9 & 120.6 & 130.2 & 126.9 & 0.49 \\
761100701* & 2015 Aug 16 & 90 & 85.4 & 85 & 87.8 & 0 \\
761101001 & 2015 Aug 30 & 139 & 119.2 & 132.9 & 128.5 & 0 \\
790380501 & 2017 Feb 1 & 143.2 & 33 & 65.5 & 54.5 & 0.5 \\
790380601 & 2017 Feb 5 & 143.2 & 86.6 & 117.92 & 100.3 & 0.25 \\
790380801 & 2017 Feb 7 & 143.2 & 101.2 & 131.9 & 114.6 & 0.25 \\
790380901 & 2017 Feb 11 & 143.2 & 118 & 137 & 133.8 & 0.5 \\
790381401 & 2017 Feb 13 & 145.7 & 112.4 & 140.1 & 132.9 & 0 \\
790381501 & 2017 Feb 15 & 145.7 & 136.6 & 139.4 & 140.2 & 0.5 \\
790381001* & 2017 Feb 21 & 97 & 77.7 & 90.5 & 93.4 & 0 \\ 
\hline
TOTAL &  & $1750\, ks $  & $1372\, ks$  & $1560\, ks$  & $1511\, ks$  &  \\
\hline \hline
\end{tabular} 
%}
\begin{flushleft}
\tablecomments{\footnotesize * PN small window  observation.}
\end{flushleft}
\end{table*}

\section{DATA PROCESSING AND SOURCE DETECTION} \label{section2}

The present XMM-Newton survey comprises 13 observations taken in 2015 and 2017 covering an area of $30\arcmin \times 30\arcmin$  centered at the blazar 1ES 1553+113  $(RA = 238^{\circ}  55\arcmin 45.48\arcsec,\, DEC = 11^{\circ} 11\arcmin 24.36\arcsec )$ \citep{Nicastro2018}. The stacked exposure time for the 13 observations is $1.75\, \mathrm{Ms}$. We processed the EPIC (PN, MOS1, and MOS2) data of our observations with the XMM-Newton Science Analysis Software version 17 \citep[SAS,][]{Gabriel2004}.  More specifically, for each EPIC observation, we used  the package \texttt{epicproc} (\texttt{epproc, emproc}) to process the data, extract images, and lightcurves. 

Observations were then filtered for periods of high background caused by soft protons as follows: first, we used the tool \texttt{evselect} to  create source-free 10-12 keV background light curves (with bin-size of 100s), for each observation and for each available instrument.  Then, we employed the task \texttt{bkgoptrate} in those light curves to identify the optimum background rate cut threshold which maximizes the signal-to-noise (S/N) ratio  for  a given background\footnote{\url{https://xmm-tools.cosmos.esa.int/external/sas/current/doc/bkgoptrate/index.html}}. We found that the 6 observations taken in 2015 show a much lower background compared to the 7 observed in 2017 (77 ks of PN high background removed in  2015 versus 296 ks in 2017). The final exposures are  $1.372,\, 1.56$, and $1.511\, \mathrm{Ms}$ for PN, MOS1, and MOS2, respectively (see Table~\ref{tabla1}). 

\subsection{Stacked Source detection} \label{section2.1}

For the process of source detection on the stacked images we used the new standardized XMM-Newton approach with the new task \texttt{edetect$\_$stack}\footnote{\url{http://xmm-tools.cosmos.esa.int/external/sas/current/doc/edetect_stack/}}, considering the same parameters and the five standard energy bands $(0.2 - 0.5,\, 0.5 - 1,\, 1 - 2,\, 2 - 4.5,\, 4.5 - 12\, \mathrm{keV})$ as in the 3XMM and 4XMM catalogs \citep{Rosen2016A,Webb2020,Traulsen2020}. The task  \texttt{edetect$\_$stack}  was prepared to perform standardized EPIC source detection on individual and  overlapping fields of observations taken at different epochs. This task includes runtime improvements and comprises twelve stages, which are run subsequently. In every stage it creates and uses in parallel data products as coupling images, exposure maps,  background maps, and detection masks for each observation, instrument, and energy band \citep[for more details, see][]{Traulsen2019}.

\subsection{Stacked source list and Maximum Likelihood Fitting} \label{section2.2}

\begin{figure}[!ht]

\centering
\plotone{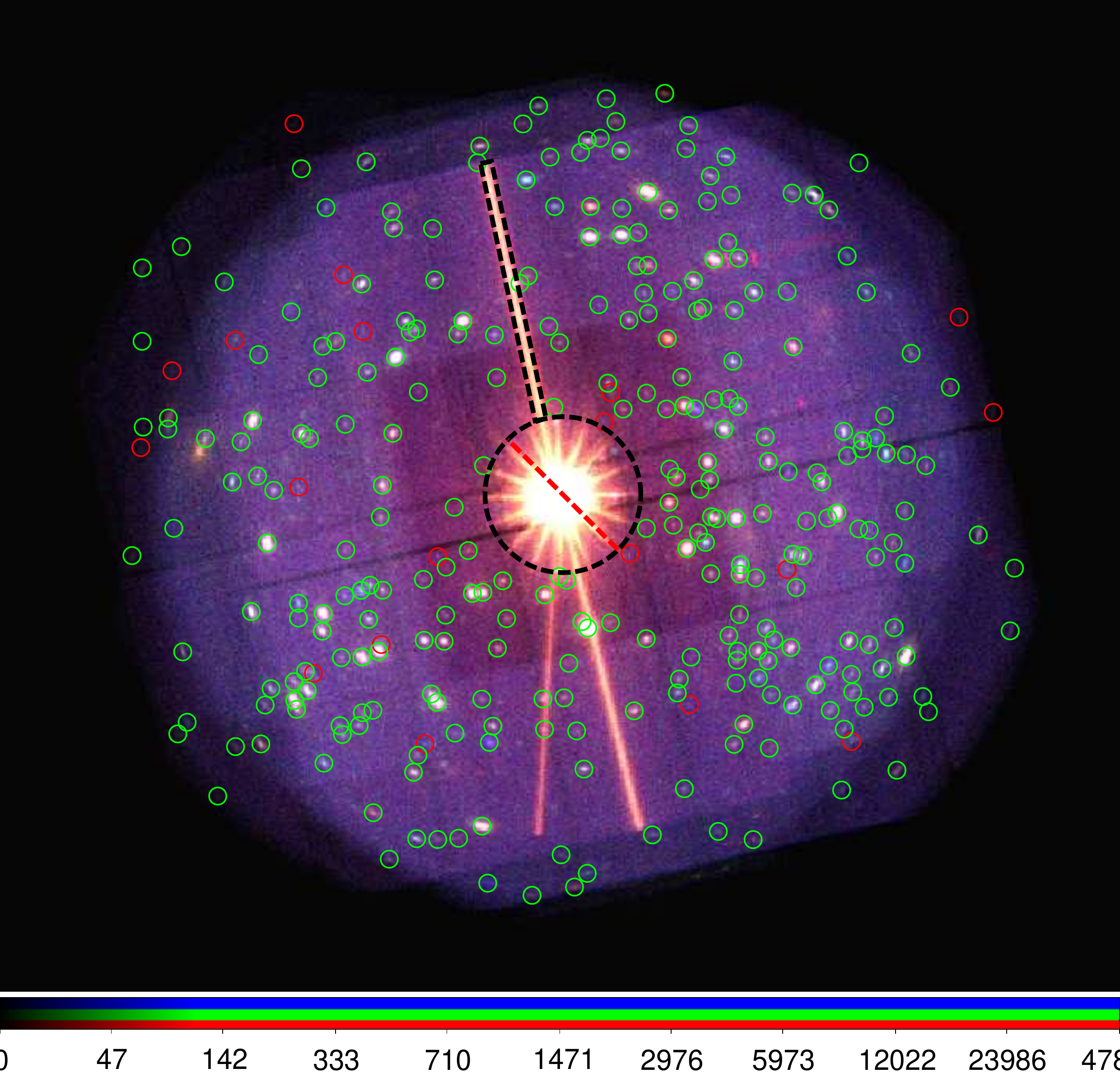}

\caption{Composite mosaic image  of all observation of the survey at 3 different bands: $0.2 - 1\, \mathrm{keV}$ (red), $1 - 2\, \mathrm{keV}$ (green) and  $2 - 12\, \mathrm{keV}$ (blue). Green circles mark the 301 X-ray point-like sources detected in our final source list. Red circles mark a subsample of 19 objects detected in only one pointing. The circular and rectangular black doted regions mask the
blazar 1ES 1553+113 and the out of time events, respectively. The color bar is in counts.}
\label{fig3}
\end{figure}

The task \texttt{edetect$\_$stack} runs \texttt{emldetect}  to calculate the X-ray source parameters of the catalog (fluxes, count rates, source counts, Maximum Likelihoods, hardness ratios) in each band for the PN, MOS1, and MOS2 cameras by fitting the instrumental PSF convolved with the source counts distribution in each energy band and camera \citep{Hasinger1993}. \texttt{emldetect} computes the likelihood significance $L$ for each source  by observation and energy band. $L$ is defined as in  \citet{Cruddace1988,Hasinger1993}

\begin{equation}
L=-ln(p)
\end{equation}

where $p$ is the probability of Poissonian random fluctuation of the counts in the detection cell, which is calculated using the incomplete Gamma function $\Gamma$  as a function of raw source counts and raw background counts in the detection box. The detection likelihoods $L_i$ for each observation are converted to Maximum Likelihoods $(ML)$, with two free parameters equivalent to perform a detection run on a single image.

\begin{equation}
ML=-ln\left( 1-\Gamma \left( \frac { \nu  }{ 2 } ,\sum _{ i=1 }^{ n }{ \frac { L_{ i } }{ 2 }  }  \right)  \right) 
\label{ML}
\end{equation}

where $\nu$ is the number of degrees of freedom of the fit ($\nu = 2 + n$, $n$ as the number of energy bands). To  minimize spurious source content, for each source, the detection likelihood is derived using the best-fit C-statistic \citep{Cash1979}, minimizing the deviation  between measured counts $c$ and the model prediction $m$ in a region of $N$ pixels.

\begin{equation}
C(c_i)=2\sum_{k=1}^{N}{m_k - c_i ln(m_k)}
\label{ci}
\end{equation}

where $c_i = c_s + c_b$ is the sum of source counts $c_s$ and background counts $c_b$ in the detection region, $m_k$ is the model prediction; as a result, $L_i$ can be characterized as $L_i= C_i  (c_i  ) - C_i  (c_b)$.  Since every observation is centered around the blazar 1ES 1553+113 ($F_{0.2-12\, keV} \simeq{2 \times 10}^{-11}\, \mathrm{erg\, s^{-1}\, cm^{-2}}$), we decided to set a circular mask of $3\arcmin$ radius centered on 1ES 1553+113  to avoid false  identifications of sources, due to the star-like pattern created by the spider structure which supports the mirrors of the telescope (see Figure~\ref{fig3}).

\subsection{Source selection process and final source list}\label{section2.3}

We set a conservative detection threshold of $ML\ge 6$ corresponding to a Poisson probability of $p\simeq 2.5 \times 10^{-3}$ (i.e. $3\sigma$). As a result, we found 483 X-ray sources, of which 49 are classified as extended. Since most of our extended sources were detected  along stray lights and along remaining residuals of the masked region around the blazar, we limited our analysis to the point-like sources.

We then proceeded to an accurate visual inspection of each source of our catalog in the final stacked image, and we were able to identify 59 additional spurious detections that appeared only in one observation and/or along stray-light strips or correspond to hot pixels, which we therefore removed from our analysis.
 
Finally, we also excluded 74 additional X-ray sources, which were detected only in the MOS1 camera and correspond to instrumental artefacts. Our final X-ray catalog contains 301 sources, all of them detected at least in the PN camera (see Appendix~\ref{AnexoA} Table~\ref{table5}), of which 6 were detected only in one band, while 38 sources in 2 bands.

\subsection{Comparisons with 4XMM catalogs} \label{section2.4}

In order to support the reliability of our results, we carried out a comparison with the most recent data release produced by XMM-Newton Survey Science Centre (SSC)\footnote{\url{http://xmmssc.irap.omp.eu/}}: the 4XMM catalogues of serendipitous sources from individual \citep[][hereafter \citetalias{Webb2020}]{Webb2020}  and overlapping \citep[][hereafter \citetalias{Traulsen2020}]{Traulsen2020} fields. Both catalogs used the same 13 observations that we presented in this paper plus 10 PN small-window calibration observations (each of $\sim 30\,ks$). The main difference between our work and theirs consists in the fact that these two catalogs were obtained by an automated process, whereas our analysis optimizes the data reduction and source detection as follows: we masked the regions affected by  the contribution of the very bright source at the center of the images and  the out of time events,  we maximized the S/N in our cleaned observations and we reduced drastically the number of spurious detections by removing bright pixels, detector features/artifacts and detections in the PSF spikes of bright sources. 

In Appendix \ref{ApendixA} we present the detailed comparison analysis of our catalog versus \citetalias{Traulsen2020} and \citetalias{Webb2020} catalogs.  Overall, we found a good consistency between both catalogs and our results:  we found 288 (\citetalias{Traulsen2020}) and 284 (\citetalias{Webb2020}) common sources within our final catalog of 301 objects.

\section{Optical/IR data set and X-ray counterparts} \label{section3}
\subsection{Optical observations and Infrared catalog} \label{section3.1}

The optical catalog used in this work was produced with observations from the OSIRIS\footnote{Optical System for Imaging and low-Intermediate-Resolution Integrated Spectroscopy (OSIRIS; \url{http://www.gtc.iac.es/instruments/osiris/}) is an imager and spectrograph for the optical wavelength range, located in the Nasmyth-B focus of GTC.
} camera at the 10m Gran Telescopio Canarias (GTC), as a result of a campaign carried out on the same XMM-Newton field \citep[PI Krongold,][]{Nicastro2018}. Four-by-four mosaic observations  were performed  with the Sloan Digital Sky Survey (SDSS) magnitude filters $u'g'r'i'z'$ centered at 350, 481.5, 641, 770.5, and 969.5 $nm$, respectively. The optical detections have signal-to-noise ratio $SNR = 3$ down to magnitude limits of 23.3, 24.9,  24.4, 23.9, and 22.7,  while for faint detections with $SNR = 2$ we used the Upper Limit (UL) of 23.7, 25.3,  24.8, 24.3, and 23.1, respectively.

The optical data were reduced with IRAF using the  \texttt{gtcmos} package \citep{Gomez2016}, while for the  source detection process we used \texttt{SExtractor}. Our analysis produced an optical catalog of $43,068$ objects; we computed their fluxes using the SDSS photometry in AB system\footnote{\url{https://www.sdss.org/dr16/algorithms/fluxcal/}}.

The infrared catalog was taken from a public repository of the Wide-field Infrared Survey Explorer (WISE) \citep{Wright2010} which additionally presents 2MASS counterparts. To cover the full XMM175UND-Field of $\approx 28\, arcmin^{2}$, we used a search cone of $20\arcmin$ radius centered on the blazar 1ES 1553+113  with the software \texttt{topcat}\footnote{\url{http://www.star.bris.ac.uk/~mbt/topcat/}}. We obtained an IR catalog of 5849 WISE sources  detected at SNR $ > 5$ in the $W1,W2,W3$, and $W4$ mid-infrared WISE bands  centered at wavelengths of 3.4, 4.6, 12, and 22 $\micron$ \citep{Cutri2013},  of which 898 sources present  2MASS counterparts in the near-infrared $J$, $H$, and $Ks$ bands. We computed the 2MASS $Ks-band$ fluxes at $2.17 \mu m$  when available, for the remaining sources we used the WISE $W1-band$ corrected by the empirical relation $Ks = 0.99 \times W1 + 0.23$ \citep{Cluver2014}.

\subsection{Optical/Infrared counterparts and Likelihood ratio technique} \label{section3.2}

\begin{figure*}
\centering
\includegraphics[scale=0.28]{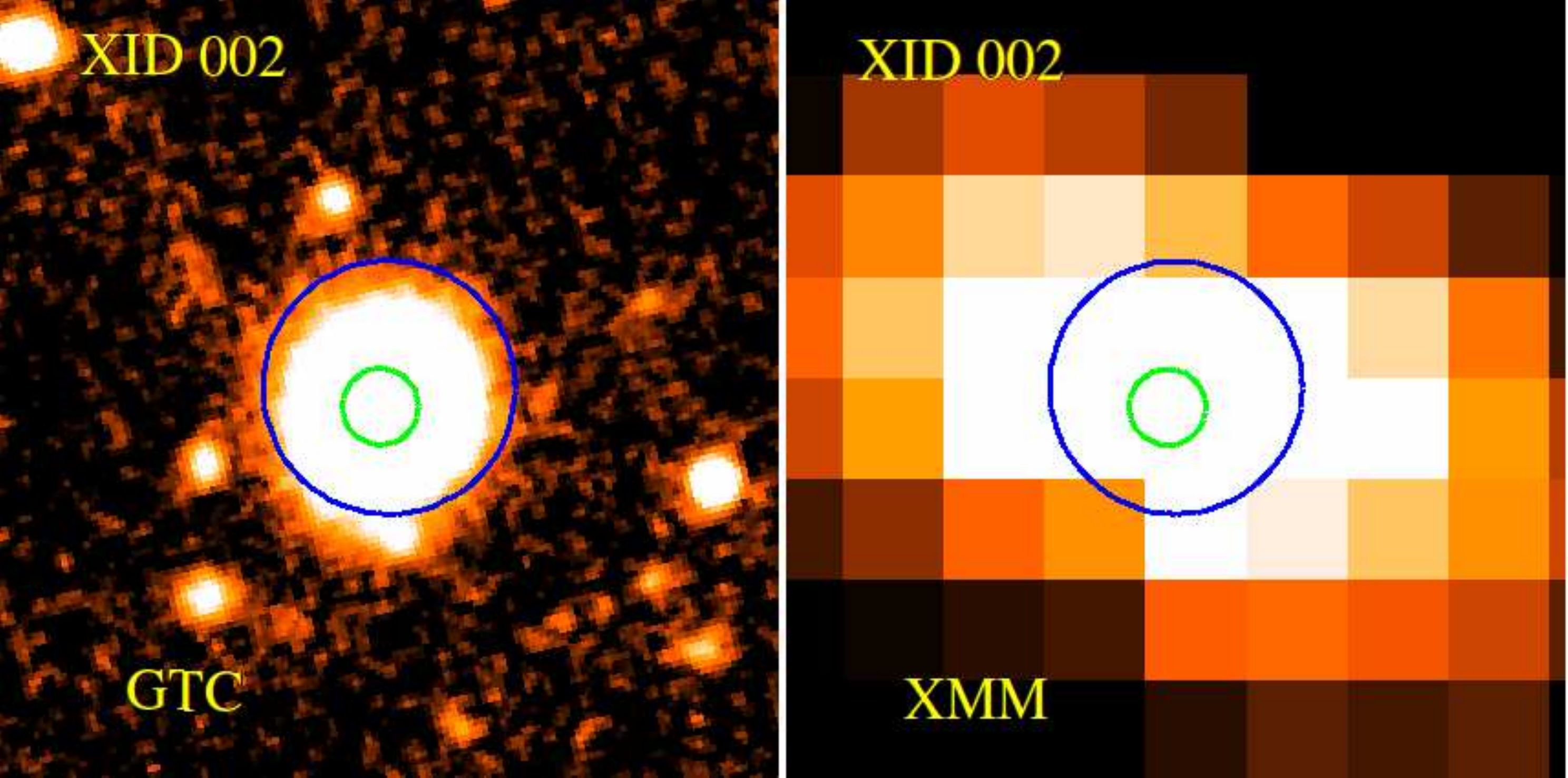} 
\includegraphics[scale=0.28]{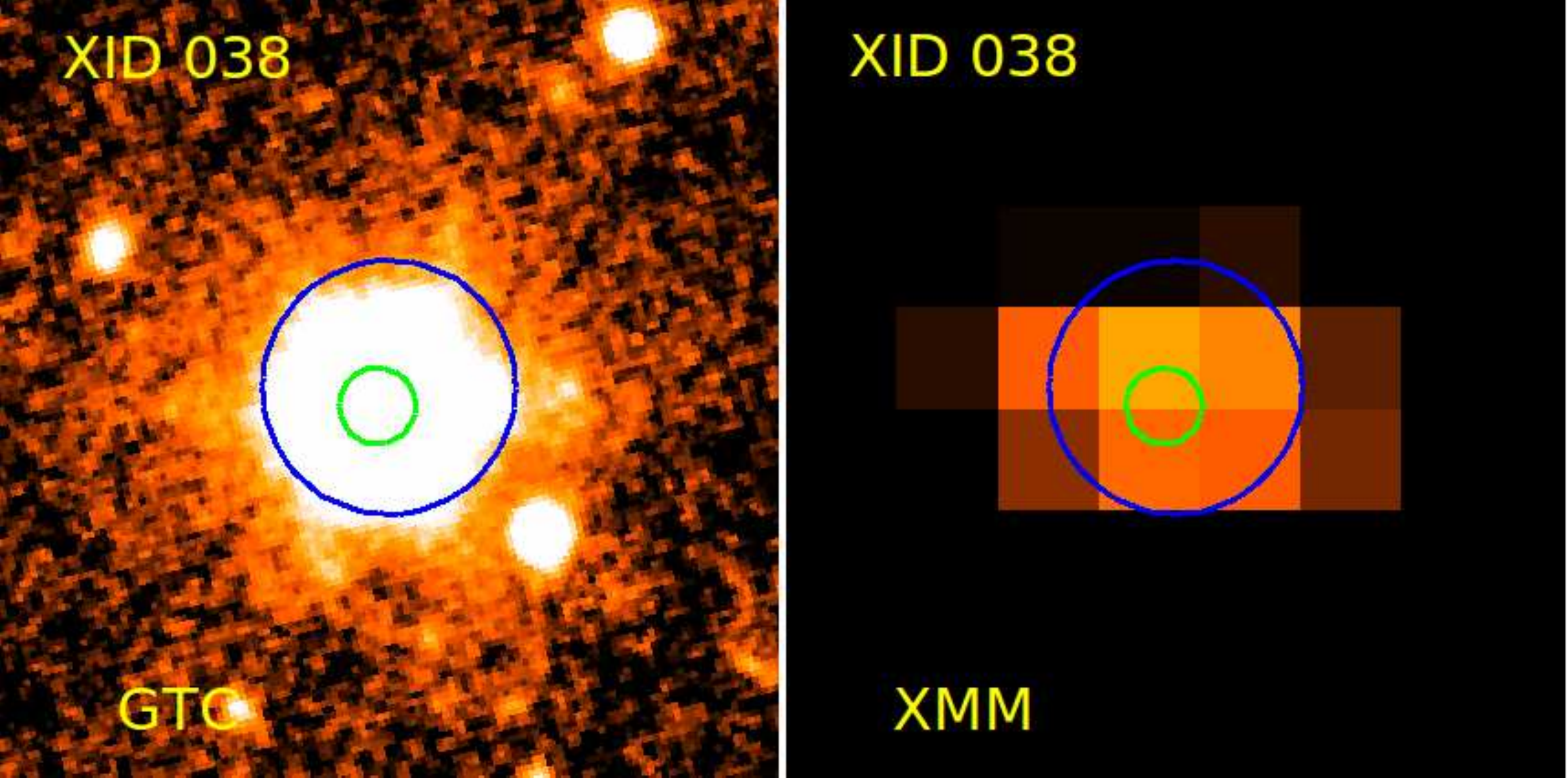} 
\includegraphics[scale=0.28]{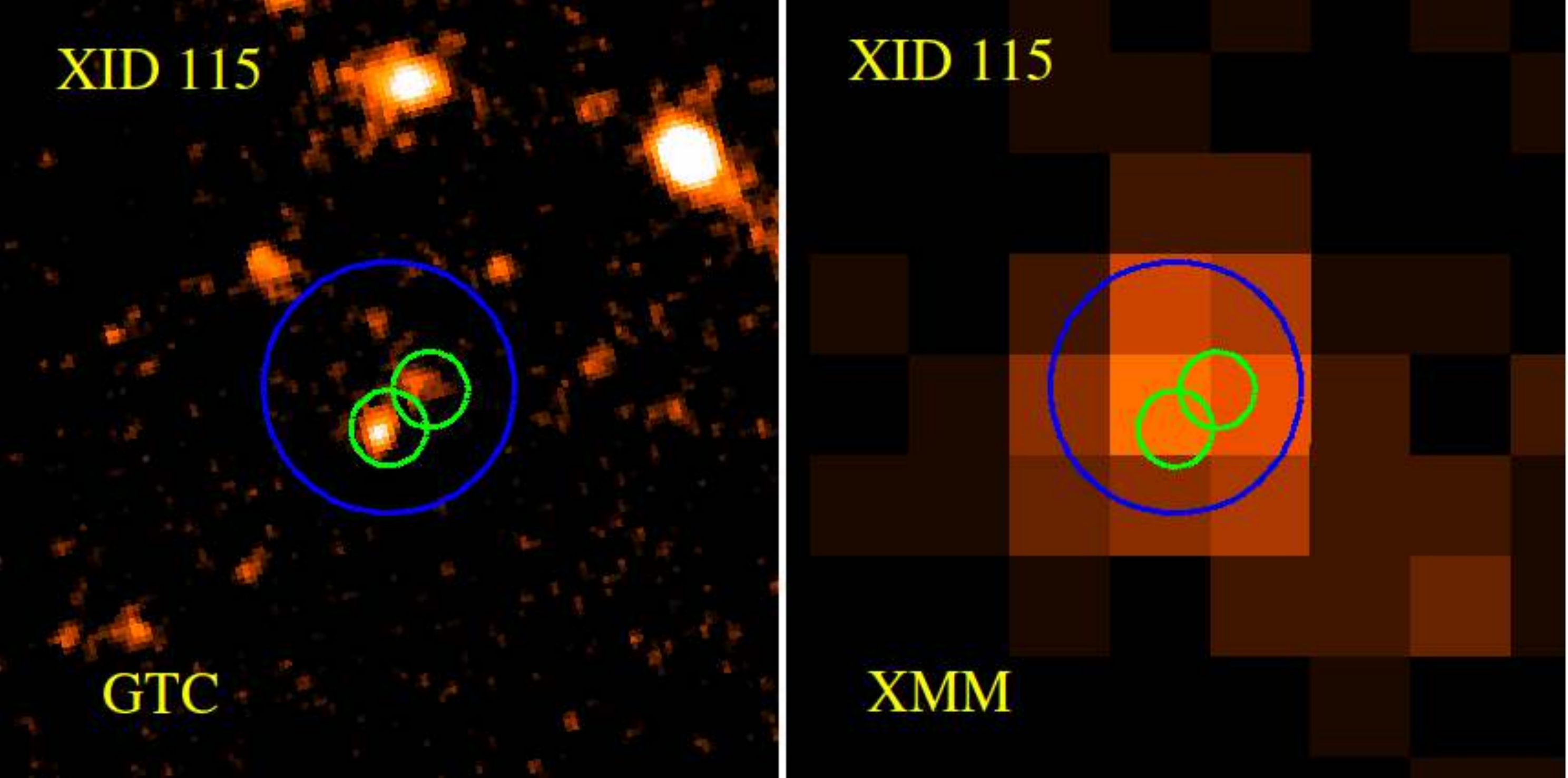} 
\includegraphics[scale=0.28]{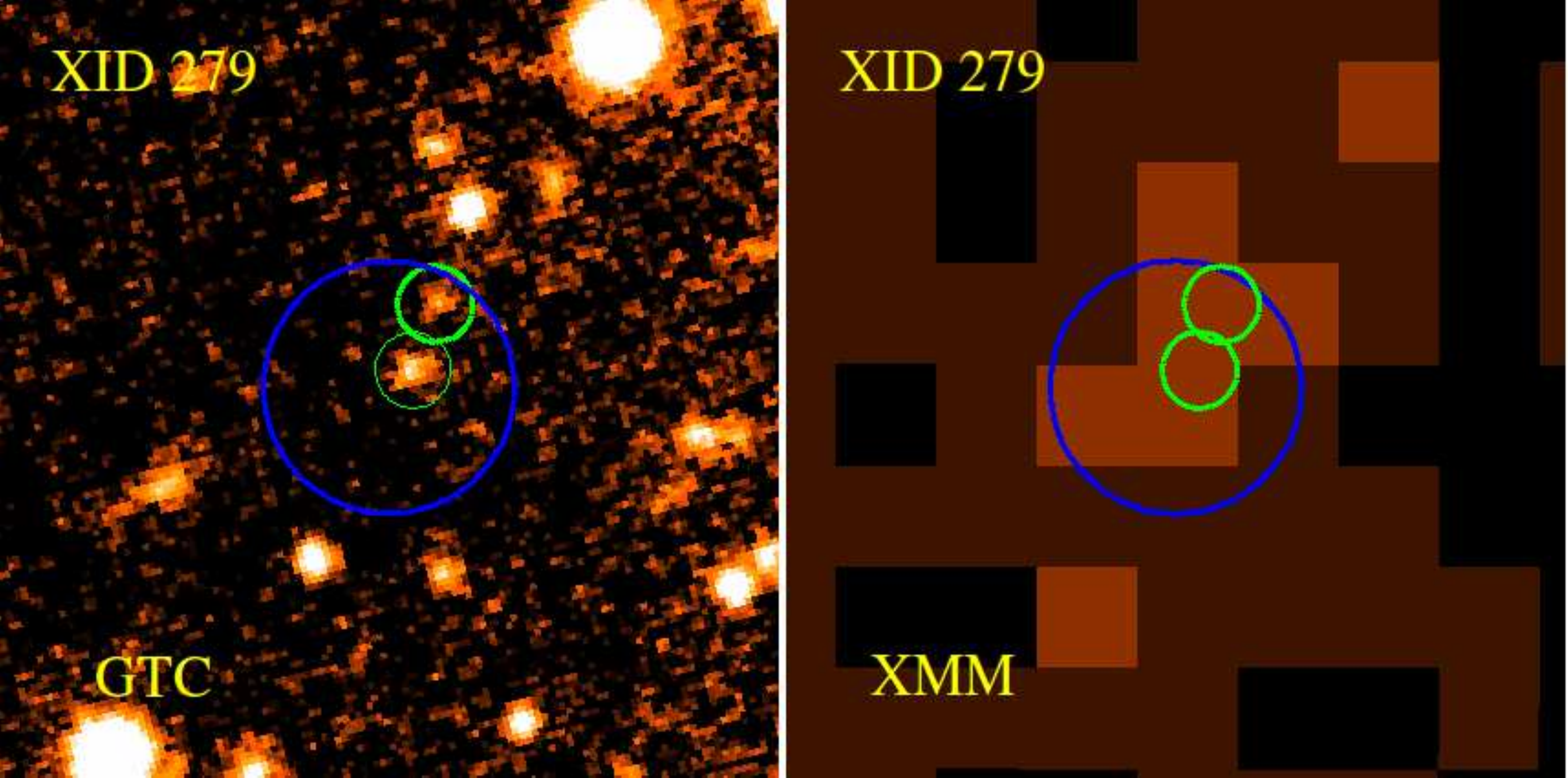}
\includegraphics[scale=0.28]{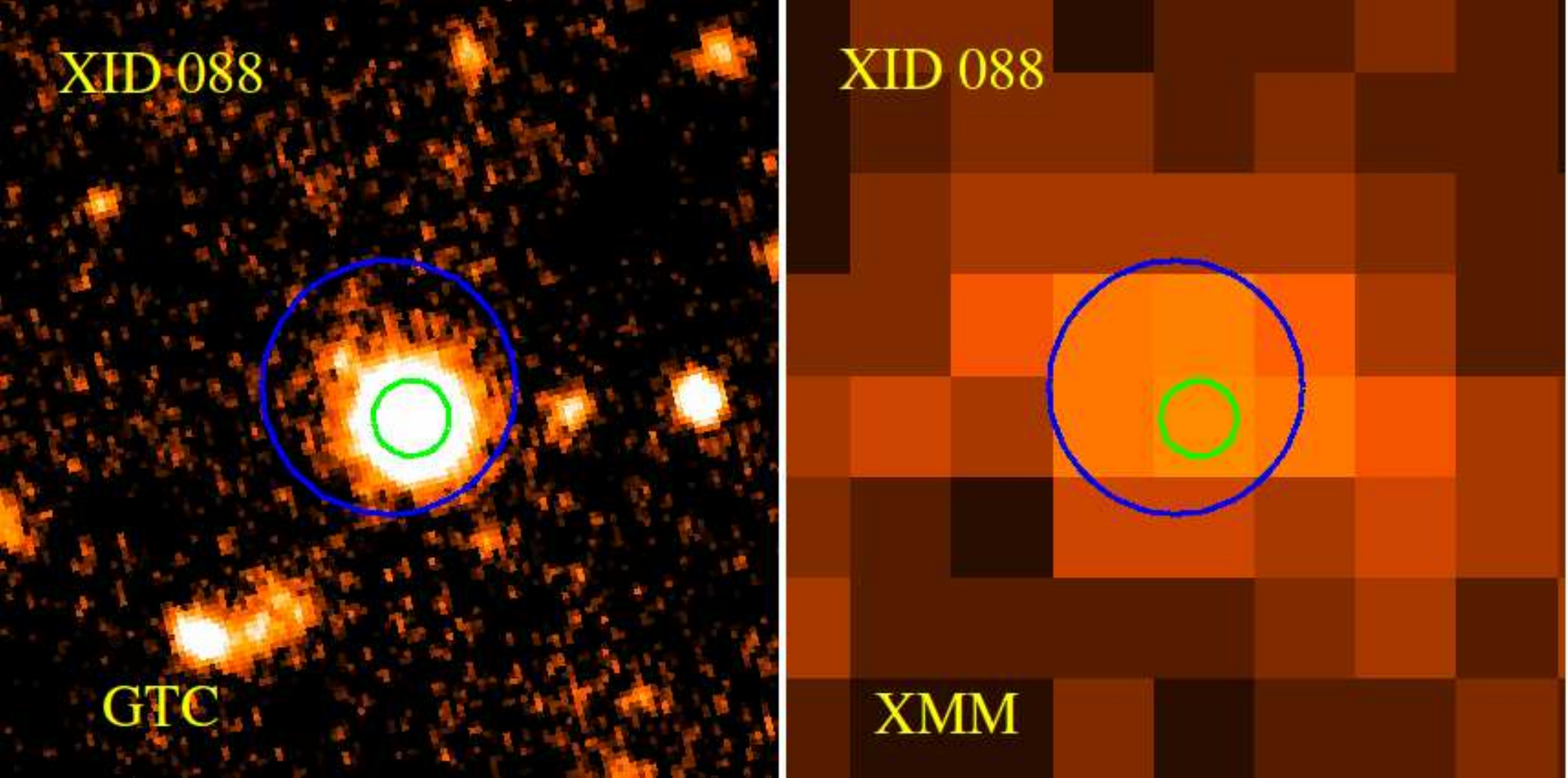}
\includegraphics[scale=0.28]{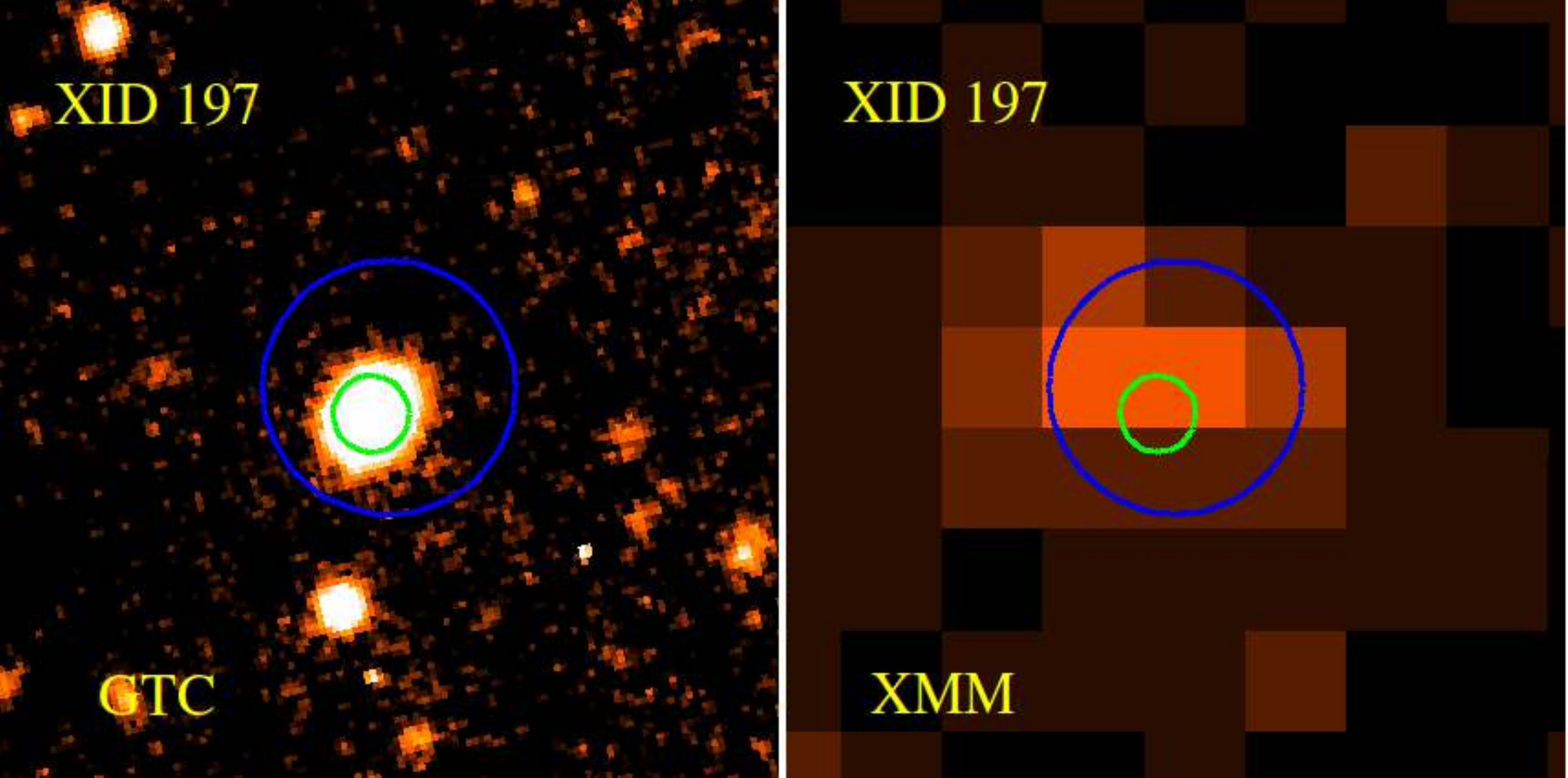} 

\caption{Example of 6 XMM-Newton randomly  selected X-ray sources with their respective GTC optical counterparts in $r'$ band images. In each chart the green circles with a radius of $1.5\arcsec$ mark the position of the optical counterpart sources, the blue circles with a radius of $5\arcsec$  are centered on the XMM-Newton  position. Every box is $0.5\arcmin \times 0.5\arcmin$ across. }
\label{OX}
\end{figure*}

The optical and infrared identifications for the X-ray sources were obtained by using the likelihood-ratio technique  \citep{Sutherland1992} considering a significance of 3$\sigma$ \citep{Brusa2007,Brusa2010,Ranalli2013,Luo2017,Chen2018}. We used the likelihood-ratio technique as described in \citet{pineau2011},  using the plugin \texttt{Xcorr} developed within Aladin. The likelihood ratio (LR) is defined as the ratio between two probability densities (see equation~\ref{LR}); first the probability to have a real association counterpart.
\begin{equation}
dp(r| true)=re^{-r^2/2}dr
\end{equation}
Second the probability that the identification is due to background fluctuations. 
\begin{equation}
dp(r| false)=2\lambda rdr
\end{equation}
Therefore LR has the expression: 
\begin{equation}
\label{LR}
LR(r)=\frac{dp(r|true)}{dp(r|false)} = \frac{e^{- r^2/2}}{2\lambda}
\end{equation}

where $r = d/\sqrt{\sigma^{2}_{X}+ \sigma^{2}_{O}}$ and $\lambda = (\sigma^{2}_{X}+ \sigma^{2}_{O}) \times N(m)$. $d$ is the angular distance that separates both sources, $\sigma_X$ (X-ray) and $\sigma_O$ (optical or infrared) are the positional error and $N(m)$ is the angular density of objects with magnitude $m$ \citep[for more details, see][]{pineau2011}.

The X-ray source positional error used in our sample is defined as: 

\begin{equation}
POSERR = \sqrt{RADEC\_ERR^{2} +  SYSERRCC^{2}}
\end{equation}

It was obtained by the quadrature combination of the  systematic positional error \textit{SYSERRCC}  due to systematic uncertainties (e.g. pointing uncertainties and cross-calibration in the stacked observations), and the statistical positional error $RADEC\_ERR$, defined as: 

\begin{equation}
    RADEC\_ERR = \sqrt{\sigma_{\alpha}^{2}+\sigma_{\delta}^{2}}
\end{equation}

Where \textit{$\sigma_{\alpha}$}  and \textit{$\sigma_{\delta}$}  are the $1\sigma$ errors on the image coordinates. For our analysis we considered a mean systematic error $SYSERRCC=0.43\arcsec$. This value was taken from \citet{Traulsen2019} who compared the position offsets of a catalog  of $71,951$ unique X-ray sources (from  $1,789$ overlapping XMM-Newton observations) and a set of associated Quasars from SDSS-DR12 \citep[see][Figure 15]{Traulsen2019}. As a result,  we obtained a mean source X-ray positional error $POSERR = 0.66''\pm 0.25$.

We found 244 X-ray sources with at least an optical or infrared counterpart association  ($81\%$ of the X-ray sources), of which 137 present both optical and IR counterparts,  90 only optical and 17 only IR counterparts (e.g. 227 optical and 154 IR counterparts). To illustrate a few examples of those objects, in Figure~\ref{OX}, we present a set of XMM-Newton and GTC (in r band) images of 6 X-ray sources with their corresponding optical counterparts.

\begin{figure}[!t]
\includegraphics[scale=0.24]{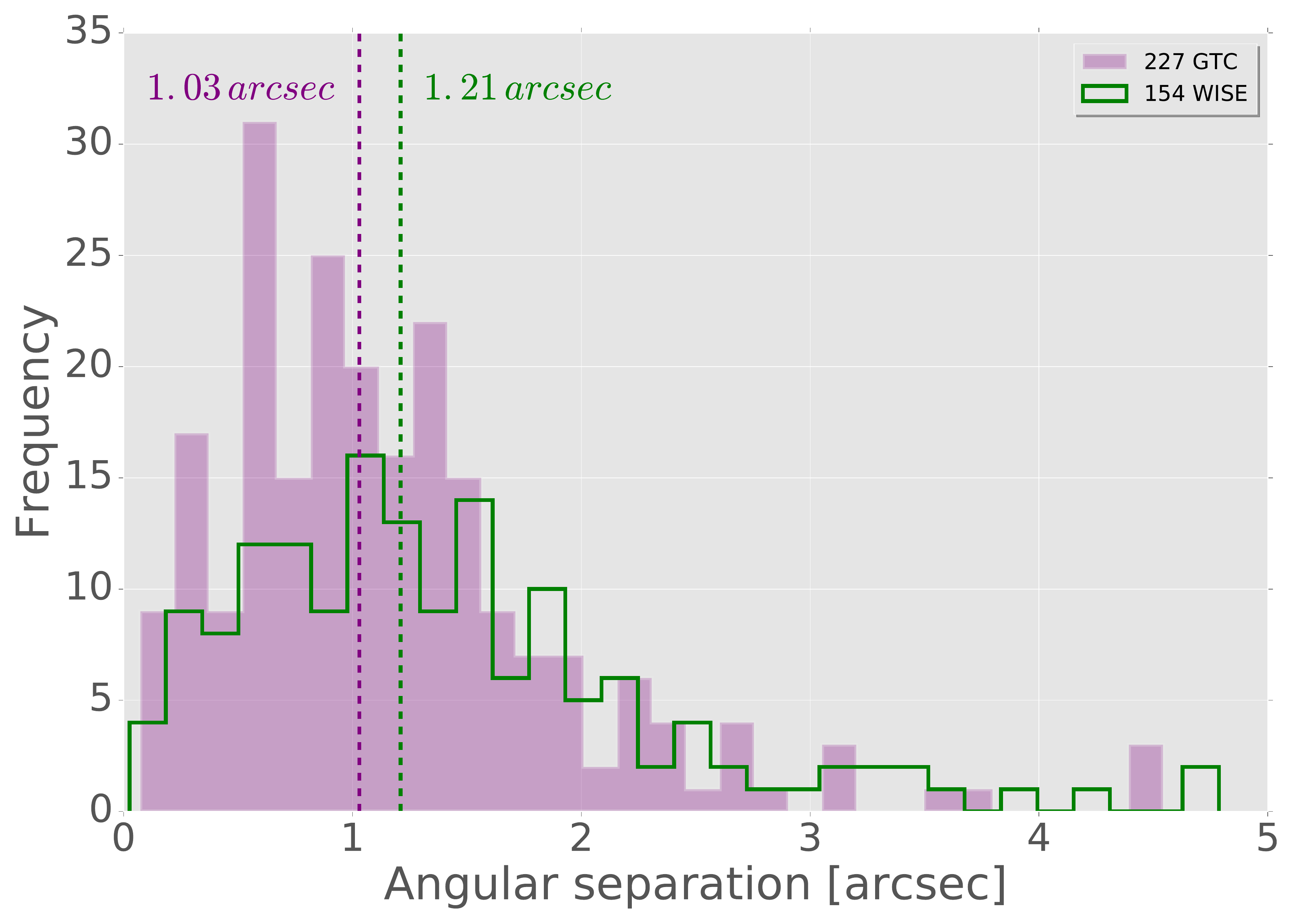} 
\caption{Angular distance distribution for GTC (227, purple filled) and WISE (154, green unfilled) counterpart candidates,  respectively, as a result from the cross-correlation procedure by \texttt{Xcorr}.}
\label{fig10_12}
\end{figure}

In Figure~\ref{fig10_12}, we plot the angular separation distribution in arcseconds, resulting from the cross-correlation of the X-ray with GTC and WISE sources, which yielded a median angular separation of $\sigma = 1.03 \arcsec$ and $\sigma = 1.21 \arcsec$, respectively. For sources which are associated with two optical counterparts (22), we considered the one with highest $LR$ for further analysis (see Appendix~\ref{AnexoA} Table~\ref{table5}).

\subsection{Spectroscopic and photometric redshift} \label{section3.3}

\begin{figure}[!ht]
\includegraphics[scale=0.33]{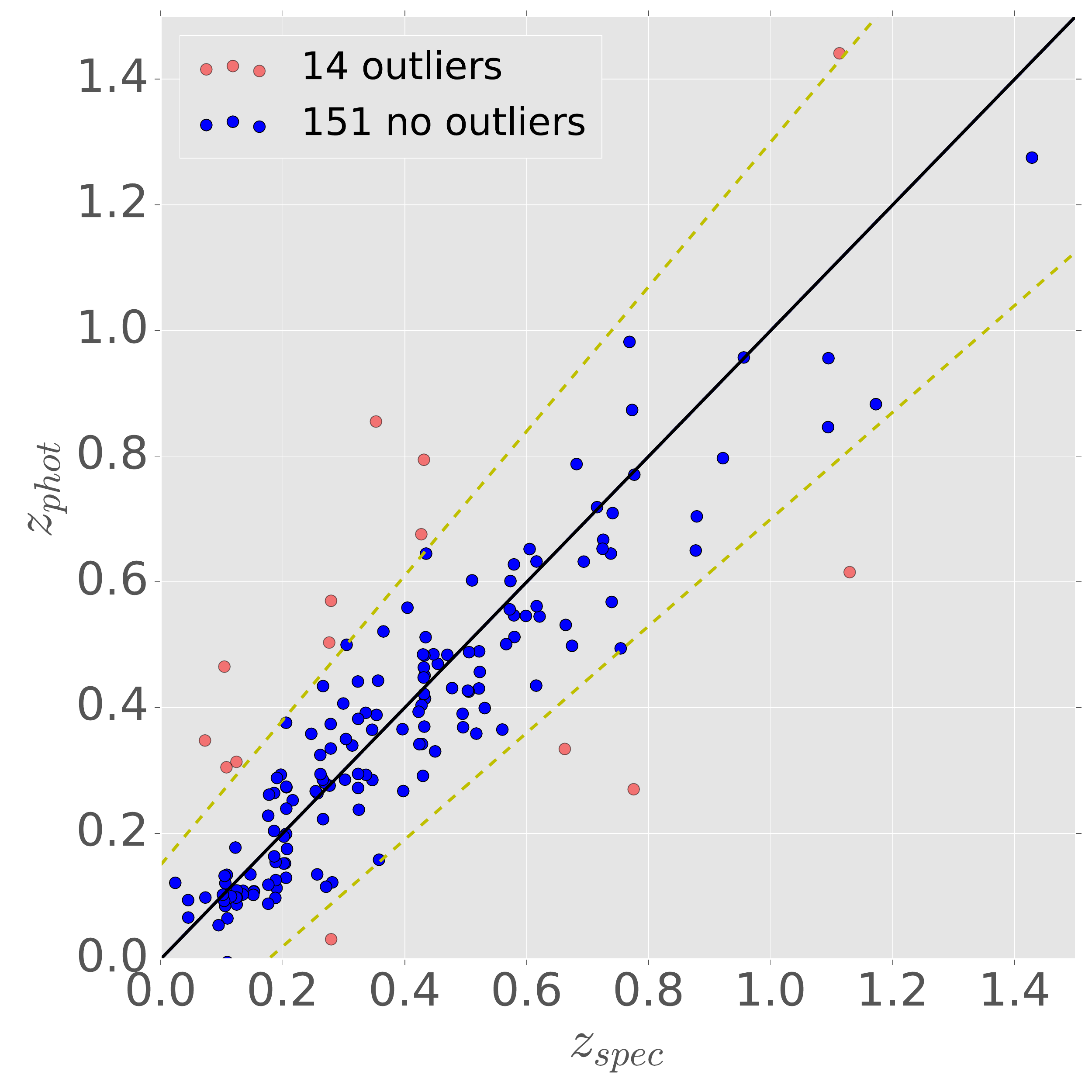} 
\includegraphics[scale=0.33]{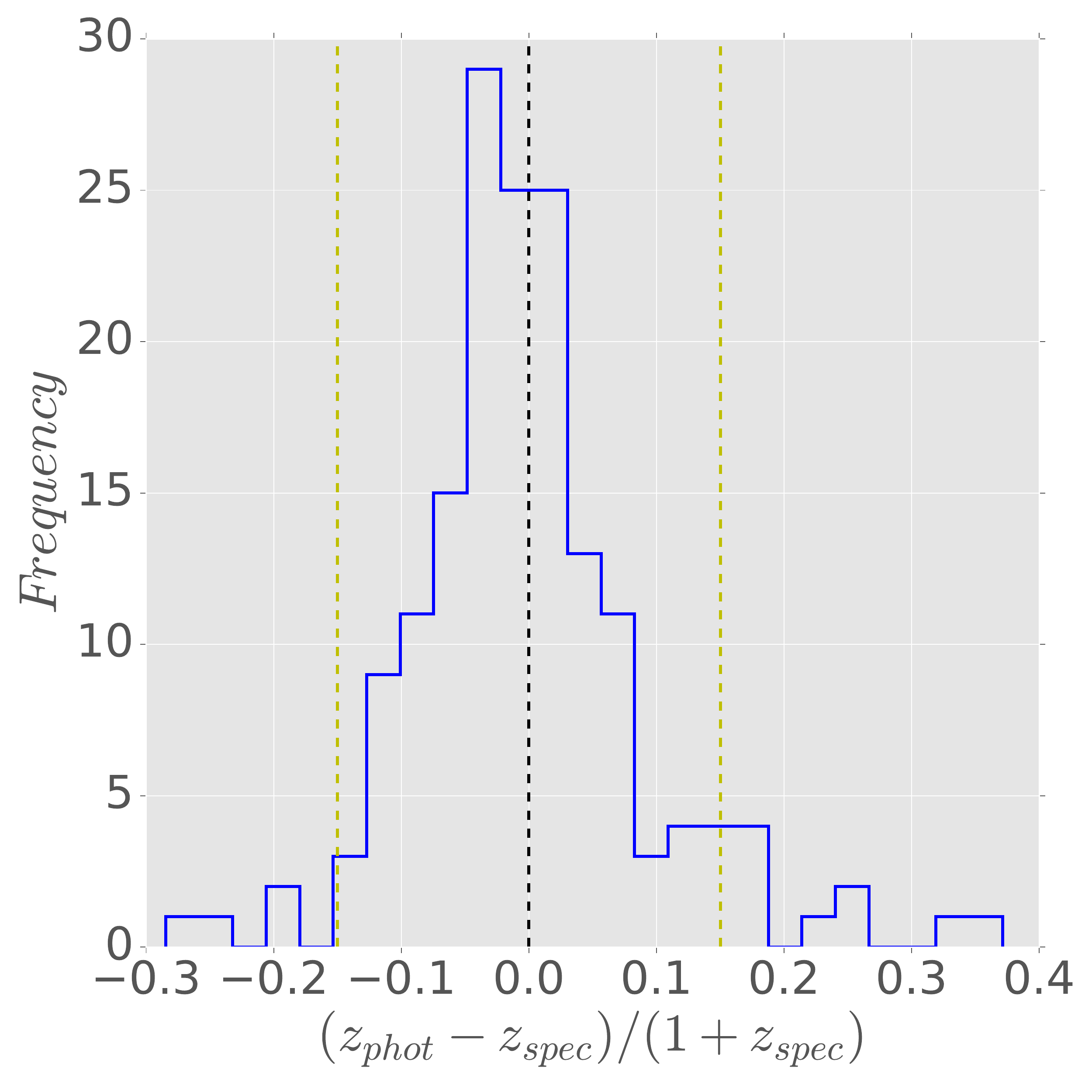}
\caption{ Upper panel: Photometric vs spectroscopic redshifts distribution for our test sample composed by 165 ($20\%$) sources from our spec-{\it z} catalog. Lower panel: residual histogram between the photo-{\it z} and the spec-{\it z}. The black solid line in both plots represents the ideal case when $z_{spec} = z_{phot} $, and the dashed yellow lines limit the confidence region $|z_{spec}-z_{phot}|<0.15 \times(1+z_{spec})$ for outliers (red solid points).}
\label{zphot}
\end{figure}

\begin{figure*}
\centering
\includegraphics[scale=0.33]{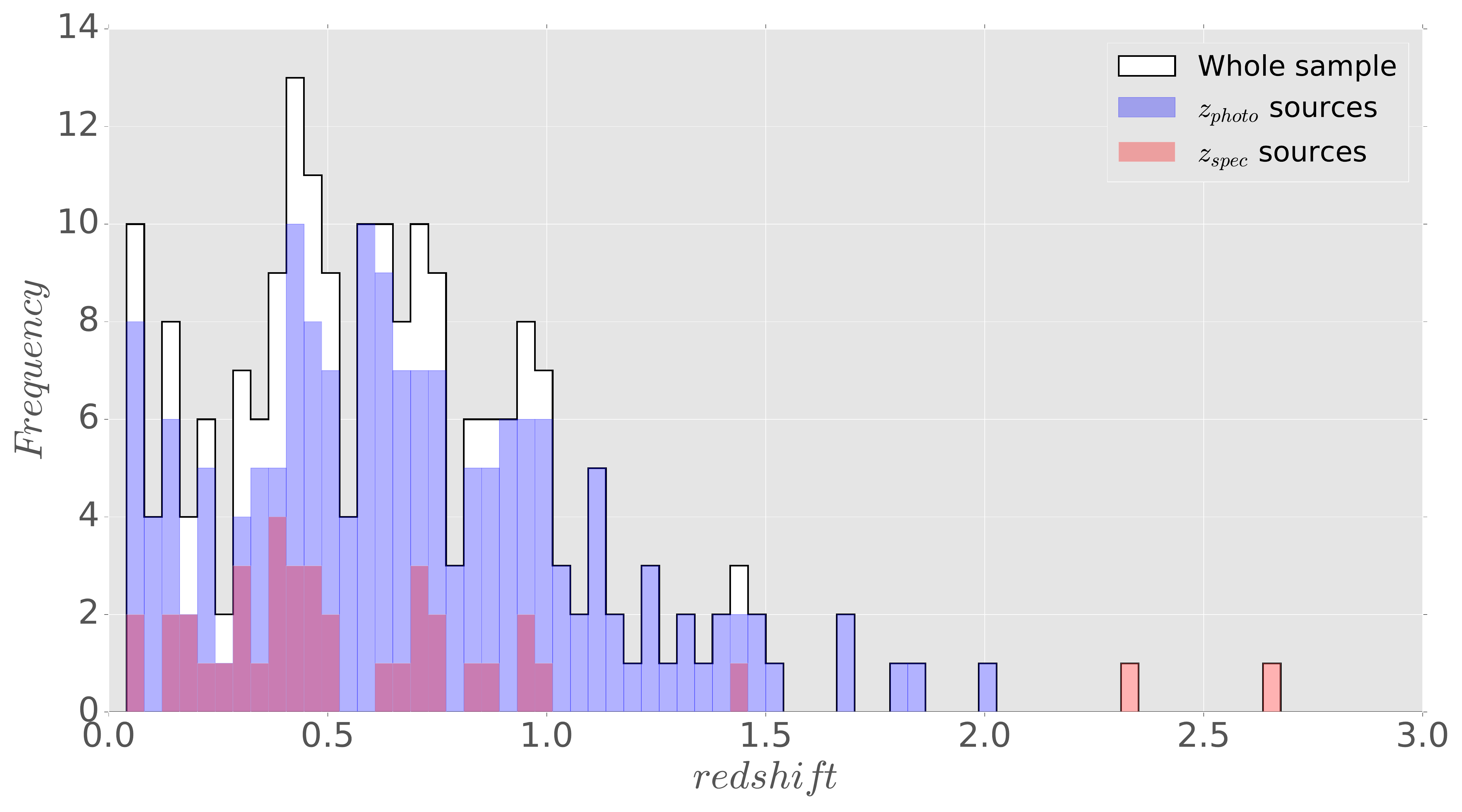} 
\caption{Redshift histogram of 211 sources of our X-ray catalog with $z_{spec}/z_{photo}$ estimations (white bars) with a bin size of 0.04. The blue and red bars respectively represent the photo-{\it z} (172) and spec-$z$ (39) sources. %The vertical dashed line marks a peak at $z \sim 0.43 \pm 0.066$. 
}
\label{histz}
\end{figure*}

The photometric redshifts were obtained with \texttt{PhotoRApToR}, a tool for photo-{\it z} calculation based on the machine learning model  MLPQNA (Multi Layer Perceptron trained by the Quasi Newton Algorithm) \citep{Cavuoti2015}.

\texttt{PhotoRApToR} uses a modern algorithm based on a neural network that was trained by using only the spectra of sources detected within our optical catalog, with the aim to execute a well-controlled experiment. The sources used for training the code present the following advantages: 1) they are observed with the same instruments, optical bands, and observing conditions; 2) they were detected in the same field (i.e. equal Galactic absorption).

An advantage of this method is that the algorithm does not require classical galaxy templates, therefore  it is not affected by its limitations (for example, some sources that may be difficult to characterize).  Moreover, thanks to the fact that most of our training set is composed by emission-line galaxies rather than absorption line objects (see Appendix \ref{gtcmos}), our code is optimized to detect and estimate the photo-$z$ of emission-line sources. Therefore, because we expect that most of our X-ray AGNs with spec-$z$ are emission-line objects, we are confident that our photo-$z$ estimations are statistically corrected.

\subsubsection{The training sample}
We used a set of 824 sources with good spectroscopic redshifts quality observed in the XMM175UND-Field; this spec-{\it z} catalog is  a combination of a recent observational campaign by \citet{Johnson2019} with 762 objects with $r' < 23.9$ magnitude limit, 29 from SDSS DR16 \citep{Ahumada2020} with $r' < 21$ and  33 sources  from our own GTC-Osiris spectroscopic observations with $r' < 23.4$. In Appendix \ref{gtcmos} we show the procedure executed for the analysis of these 33 optical spectra.

 Additionally, 39 sources of our X-ray catalog have an optical counterpart association with our spec-{\it z} catalog. 32 of them were included in the training set template to estimate the photo-{\it z} and the remaining 7 in the test set, which  may improve our results in terms of accuracy. 
We trained the neural network by using the 5 optical bands $u'g'r'i'z'$ as the input parameters with an  $80\%$ (659) of our spec-$z$ catalog as the training set, leaving the remaining $20\%$ (165) to test our results (see Figure \ref{zphot}). 
Since $99\%$ of the sources of our spec-z catalog have redshifts in the $zspec = 0 - 1.5$ range, we could  constrain our photo-z in a reliable way up to $z\sim 1.5$. 

Then, we used the normalized median absolute deviation (NMAD) defined as $\sigma_{NMAD} = 1.4826\times Median(\Delta z/(1+z_{spec}))$  as an indicator of the quality of our photo-{\it z} estimation, where $\Delta z = |z_{spec}-z_{phot}|$.  We found an accuracy of  $\sigma_{NMAD}  = 0.062$ with $\sim 8.5\%\, (14)$ of outliers (e.g. $|z_{spec}-z_{phot}| > 0.15 \times (1+z_{spec})$) and a normalized standard deviation $\sigma_{\Delta z/(1+z)}$ (or $\sigma_{norm}$) of 0.064. Our results are comparable with previous works such as  the XMM-Newton survey in the COSMOS field \citep[$\sigma_{norm} = 0.05$, outliers $\approx 8\%$,][]{Brusa2007,Brusa2010}, and  the XMM-SERVS survey  \citep[$\sigma_{NMAD} = 0.040$, outliers = $8.7\%$,][]{Chen2018}.

To test the effect of the number of added X-ray sources to the training set  ($82\%$ in the analysis above), we trained our neural network considering two additional cases: using a training set without X-ray objects and with $50\%$  of them (20 out of 39). We found a fraction of  $8.4\%$ of outliers with $\sigma_{NMAD} = 0.0634$ for the first case. For the second case we found $8.6\%$ of outliers with $\sigma_{NMAD} = 0.0625$. These results show that there is no dependence with the number of X-ray sources used in the training set. We stress again that this is because most of our spectroscopic sample consists of  objects with emission lines.

After performing our photo-{\it z} calculation and considering the spectroscopic sample, we achieved a $\sim 93\%$ (211 out of 227) of redshift completeness  for our  X-ray sources with optical counterparts.  In Figure \ref{histz}, we show the histograms for our  photometric (red), spectroscopic (blue) and full (white) redshift distribution of our X-ray sources in the XMM175UND-Field.

\section{X-RAY SOURCE PROPERTIES} \label{section4}
\subsection{X-ray flux distribution}  \label{section4.1}

The final X-ray catalog consists of 301 objects, of which 87  are  detected only in the soft bands $(0.2-0.5, \,  0.5-1, \, 1-2\, \mathrm{keV})$,  17 only in the hard bands $(2-4.5, \, 4.5-12\, \mathrm{keV} )$ and 197 are detected simultaneously in both soft and hard bands. These 197 objects are  defined  as ``sources detected in the Full band $(0.2 - 12\, keV)$'' (see Table \ref{table2}).

Similar to other XMM-Newton catalogs created by the SSC \citep{Rosen2016A,Traulsen2019,Traulsen2020,Webb2020} we estimated our source fluxes using the same count-to-flux conversion factors adopted by  \citet{Mateos2009}. The model assumes a power-law spectrum with photon index $\Gamma = 1.7$ and Galactic absorption of $N_H = 3 \times 10^{20} \, \mathrm{cm}^{-2}$ comparable with the Galactic absorption of $N_H = 3.56 \times 10^{20} \, \mathrm{cm}^{-2}$ for this field. For simplicity, the fluxes for observation and energy band are obtained by using only the PN camera. The flux for each source per energy band is the average PN flux of the overlapping observations. Here we did not apply any further correction for possible individual intrinsic absorption.

To test the effect of a steeper photon index on the flux estimate, we followed two different approaches. First, we selected the 26 brightest sources (with more than 500 counts in $0.2-10\, keV$) detected in all the 13 observations where direct spectral analysis is possible. We modeled their spectra with a power law absorbed by the Galactic column density and a fixed photon index. Fluxes obtained using a photon index of $\Gamma = 1.4$ were compared  with the values obtained with a  $\Gamma = 1.7$. The second approach consists in computing the fluxes directly from their count rates by using the energy conversion  factor  from  the  XMM-Newton  User Handbook online page\footnote{\url{https://xmm-tools.cosmos.esa.int/external/xmm_user_support/documentation/uhb/epicfluxtocr.html}}. These tests show a moderate underprediction  of $17\%$ in the soft band and an overprediction of $30\%$ in the hard band. The combination of such variations is consistent with a difference of $20\%$ in the full band, in agreement with the $15\%$ reported by \citet{Mateos2009}. 

%The combination of such variations represents a higher fluxes with a difference of $20\%$ in the full band, which is consistent with the $15\%$  reported by \citet{Mateos2009}. 

%Furthermore, in Appendix \ref{ApendixA} we showed that our flux measurements are consistent with the reported by \citet{Traulsen2020} and \citet{Webb2020} with the 4XMM-DR9s and 4XMM-DR9 catalogs, respectively.

\begin{figure}[!t]
\includegraphics[scale=.28]{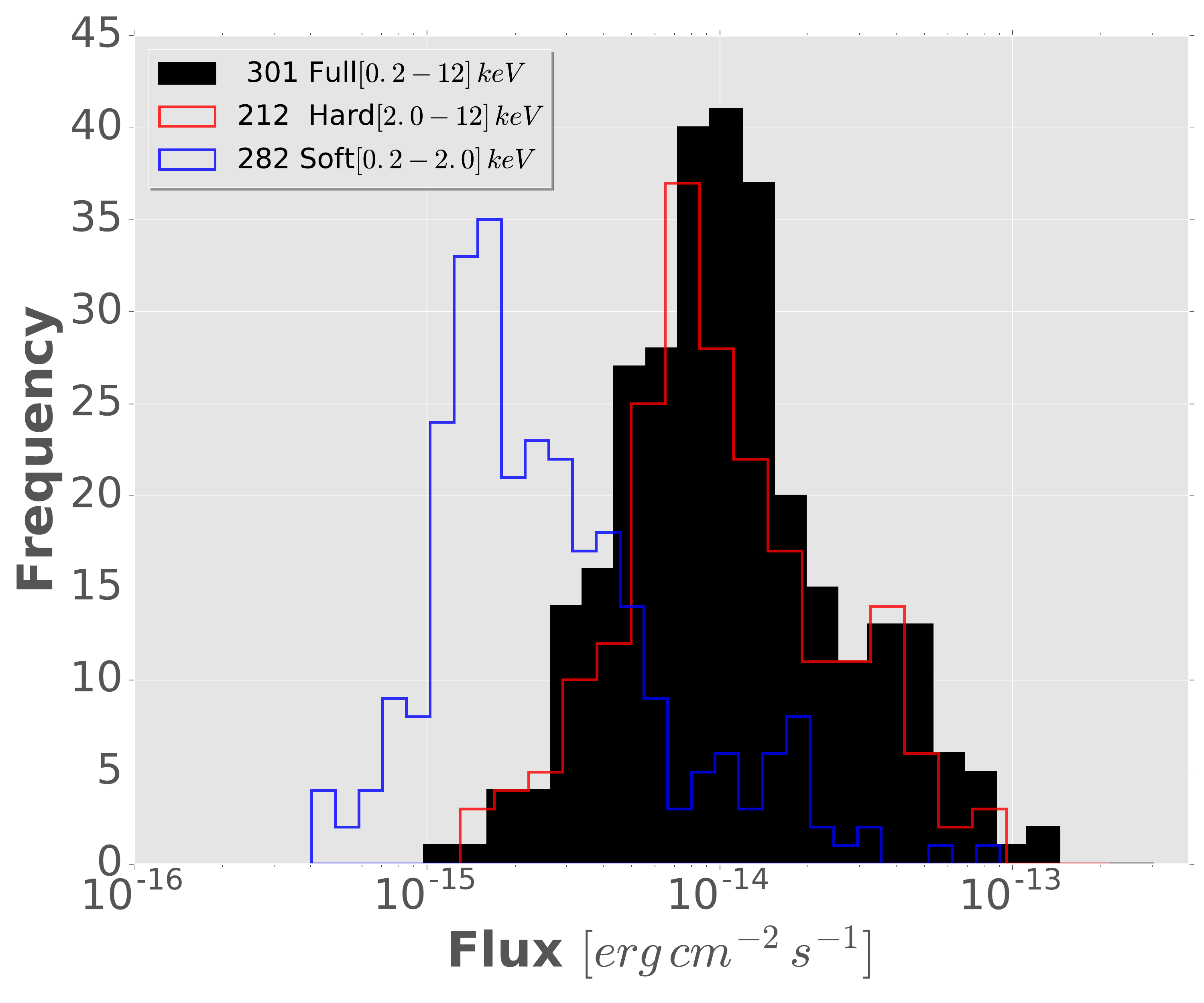} 
\caption{Flux distribution for our 301 X-ray point source catalog  in soft $[0.2 - 2\, \mathrm{keV}]$ (blue histogram), hard $[2 - 12\, \mathrm{keV}]$ (red histogram), and full $[0.2 - 12\, \mathrm{keV}]$ (black filled histogram) bands.}
\label{fig4}
\end{figure}

The flux distribution and the sensitivity limit of our survey are presented in Figure~\ref{fig4}  with the faintest sources at $4.03 \times 10^{-16}\, erg\, s^{-1}\, cm^{-2} $  in the $0.2 - 2\, \mathrm{keV}$ band, $1.3 \times 10^{-15}\, erg\, s^{-1}\, cm^{-2}$ in the $2 - 12\, \mathrm{keV}$ band and $9.8 \times 10^{-16}\, erg\, s^{-1}\, cm^{-2}$   in the $0.2 - 12\, \mathrm{keV}$ band (see Table \ref{table2}). Similar to \citet{Ranalli2013}, we considered the lowest fluxes in each band as the flux limits of our survey. Additionally, this choice is consistent with the values from the lowest sky-coverage fluxes computed from our sensitivity maps in section \ref{section4.3}.

Considering the level of background and the same spectral assumptions of both surveys, we conclude that our results are comparable with those of  \citet{Ranalli2013} for the XMM-Newton survey in the Chandra Deep Field South that presents a similar sky area of $830 \, arcmin^{2}$  (equivalent to $28.8\arcmin \times 28.8\arcmin$) but twice nominal exposure time of 3.45 Ms. In fact, they achieved an X-ray sensitivity  of $6.6\times 10^{-16}\, \mathrm{erg\, cm^{-2}\,  s^{-1}}$, in the $2 - 10$~keV band, which is roughly twice more sensitive than our survey.

\subsection[]{Sky coverage and $Log\, N(>S) - Log\, S$ analysis} \label{section4.3}

\subsubsection{Sky coverage} \label{section4.3.1}

\begin{figure}[!ht]
\centering
\includegraphics[scale=0.24]{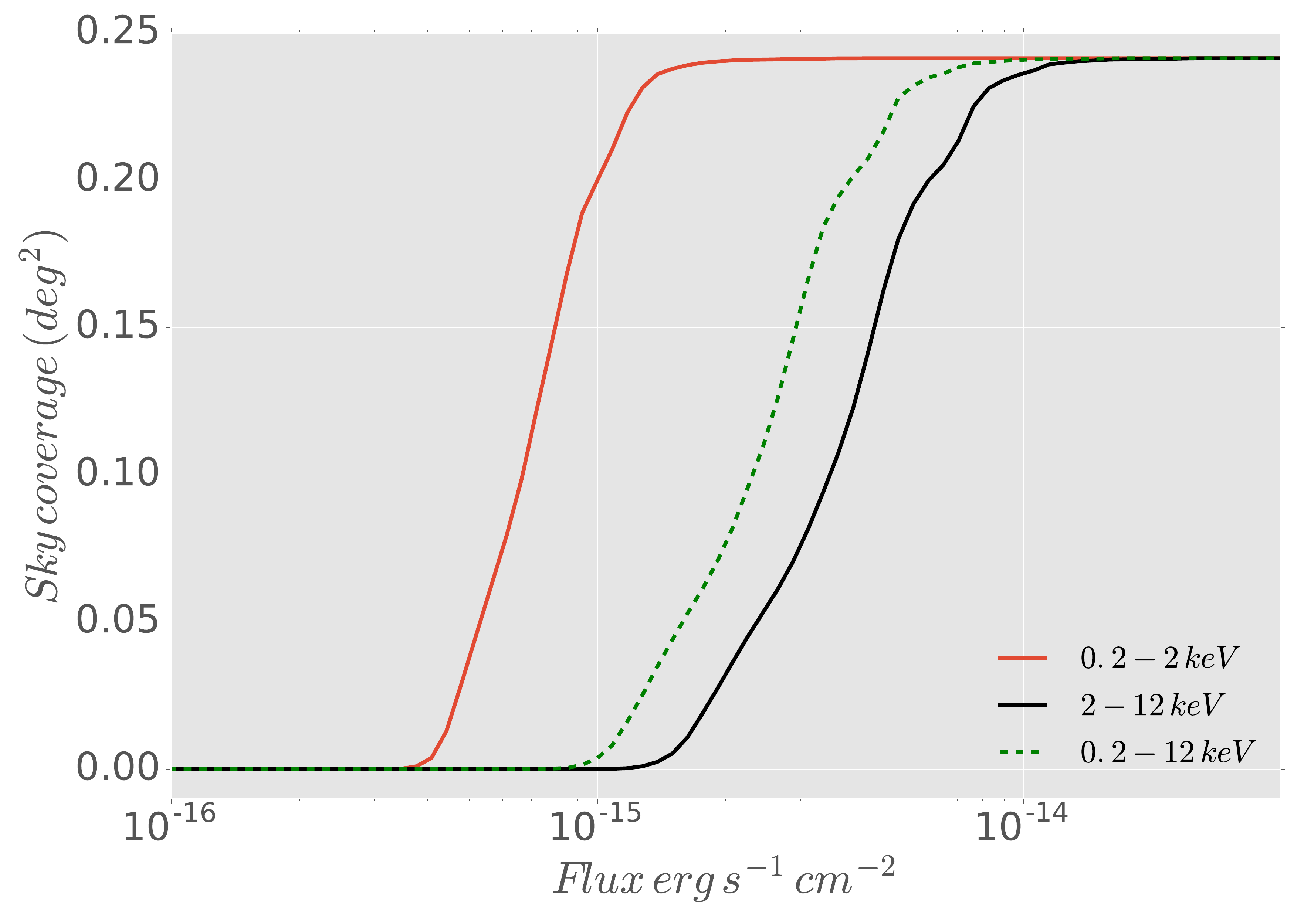}
\caption{Sky coverage curves as a function of flux in the hard (black line), soft (red line), and full (green dotted line) X-ray bands computed from the combination of the individual sensitivity maps of each observation}.
\label{sensitivity}
\end{figure}

\begin{figure*}
\centering
\includegraphics[scale=0.28]{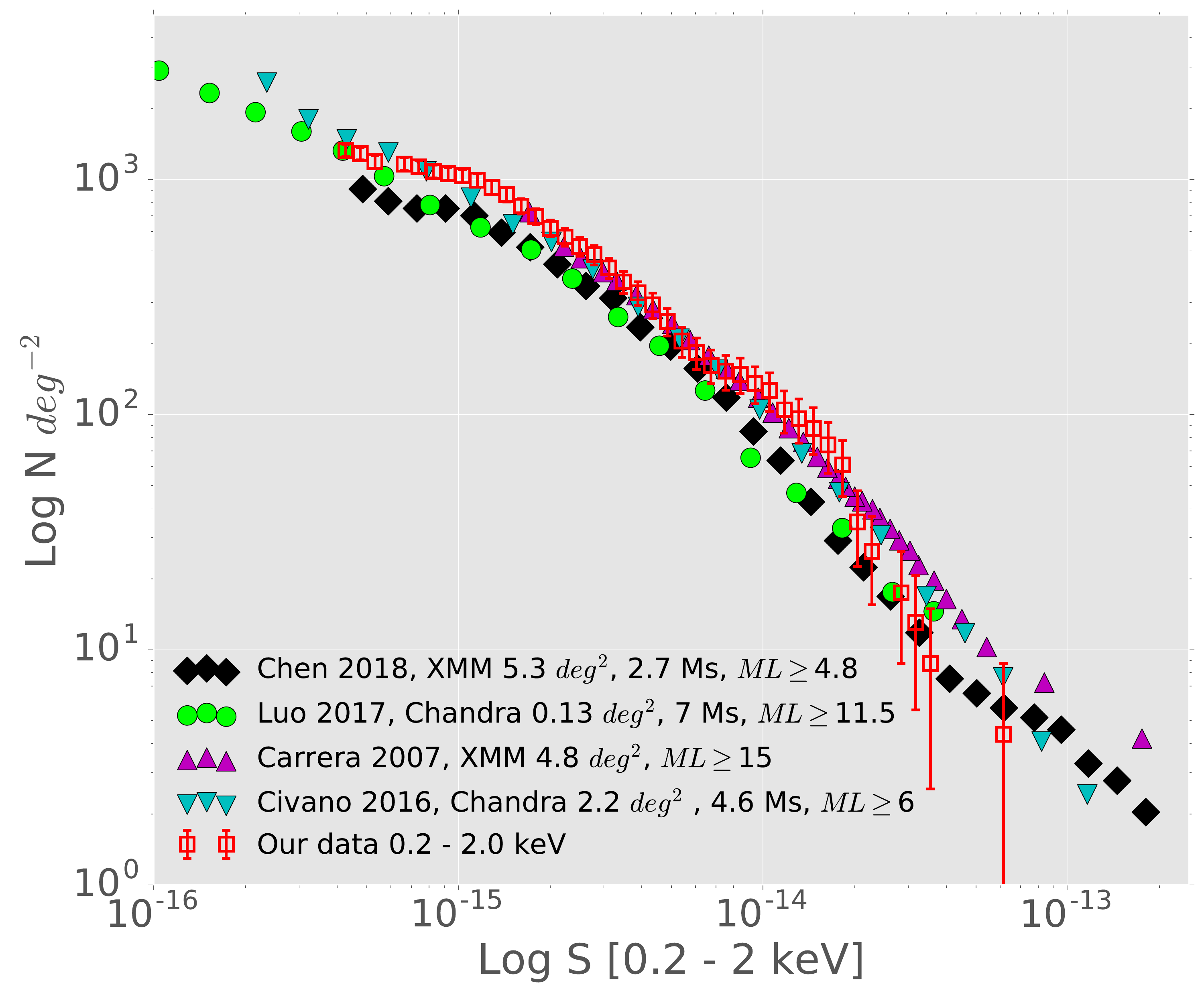} 
\includegraphics[scale=0.28]{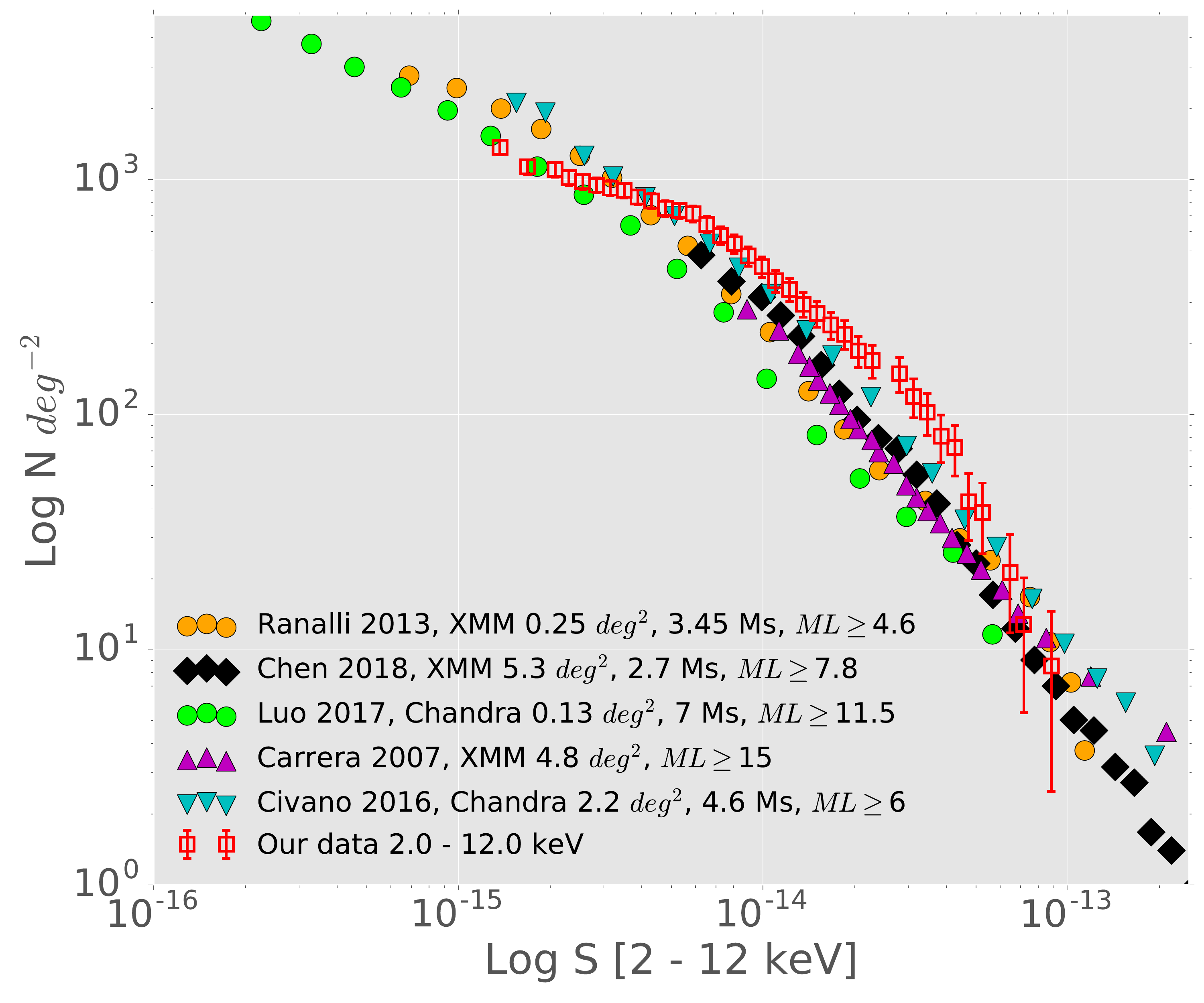}
\caption{Comparison of our $Log\, N(>S) - Log\, S$ distribution (red squares) for the 269 filtered sources observed in the soft band (Left panel)  and 205 sources observed in the hard band (Right panel) with previous representative surveys  at small $(\le 1\, deg^{2})$ and medium sky coverages $(\le 5.5\, deg^{2})$.}
\label{logNlogS}
\end{figure*}

To estimate the expected source number distribution for our survey, we proceeded as follows: first, we calculated the sky coverage  as a function of the X-ray flux from our sensitivity maps in  every observation (PN images) and energy bands. The sky coverage is defined as the solid angle within which a source with a certain X-ray flux can be detected with $ML \geq 6$. The sensitivity maps were created by the task \texttt{esensmap} during the source detection processes. Those maps represent the count rate that a source needs in order to be detected, in a specific position and energy band. These maps are produced with the same  detection threshold adopted in the source detection procedure ($ML \ge 6$, see Section \ref{section2.2}). Each map was divided into circular areas of 3 pixels radius by considering the XMM-Newton PSF size and the source detection cell of $5 \times 5$ pixels used during the source detection process (with image binning of $4\arcsec$ pixel side). We obtained  and added the corresponding count-rate and solid angle of every circular region, to obtain the cumulative survey area as a function of the mean flux limit. 

We note that the total sky coverage of the survey  is reduced to $29.5\arcmin \times 29.5\arcmin$ equivalent to $0.241\, deg^{2}$  due to the masking applied (see Figure \ref{fig3} in Section \ref{section2.3}). In Figure~\ref{sensitivity} we present the sky coverage fluxes of our survey, computed from our sensitivity maps with the lowest fluxes at $2.7 \times 10^{-16}\, erg\, s^{-1}\, cm^{-2}$, $1 \times 10^{-15}\, erg\, s^{-1}\, cm^{-2}$,  and $7.3 \times 10^{-16}\, erg\, s^{-1}\, cm^{-2}$ in the soft, hard and full band, respectively. The faintest sources detected in our catalog are consistent with these fluxes.

\begin{table*}
\caption{Summary of X-ray source counts by energy band in our XMM175UNDF catalog. The first 5 bands are the standard detection bands \citep{Rosen2016A}, the following 2 bands are for sources detected in the soft and hard bands, respectively.}
\label{table2}
%\resizebox{\linewidth}{!}{%
\centering
\begin{tabular}{cccccc}
\hline \hline 
Band & $N_{tot}^a$ & $N_{tot,obs}^b$ & $N_{fil}^c$ & $N_{fil,obs}^d$ & $S_{min}/S_{max}^e$ \\
$(keV)$ &  &  &  &  & $(10^{-15})$ cgs \\ \hline
0.2-0.5 & 148 & 0.07 & 148 & 0.07 & 0.16/14.75 \\
0.5-1 & 212 & 0.13 & 205 & 0.12 & 0.20/30.97 \\
1 - 2 & 262 & 0.26 & 251 & 0.27 & 0.25/45.11 \\
2-4.5 & 212 & 0.35 & 205 & 0.33 & 0.76/68.37 \\
4.5 - 12 & 82 & 0.35 & 81 & 0.35 & 3.63/145.59 \\
Soft & 282 & 0.25 & 269 & 0.26 & 0.4/90.45 \\
Hard & 212 & 0.35 & 205 & 0.33 & 1.3/213.96 \\
Full-band$^f$ & 197 & 0.3 & 196 & 0.3 & 0.98/304.41 \\
Only-Soft$^g$ & 87 & - & 75 & - & 0.4/5.91 \\
Only-Hard$^h$ & 17 & - & 11 & - & 3.43/36.7 \\
Full-Survey$^i$ & 301 & 0.3 & 282 & 0.29 & 0.98/304.41 \\ \hline \hline
\end{tabular}

%}
\begin{flushleft}
\tablecomments{\\\footnotesize 
$^a$ Total sources detected by band.
\\
$^b$ Fraction of obscured sources $(HR \ge -0.2)$ (see Section~\ref{section4.2}) of  $N_{tot}$.
\\
$^c$ Final number of filtered sources to compute the $Log\, N(>S) - Log\, S$ distribution.
\\
$^d$ Fraction of obscured filtered sources  (see Sections~\ref{section4.3.1} and \ref{section4.3.2} for details).
\\ 
$^e$ Minimum and maximum fluxes per band in $erg\, cm^{-2}\, s^{-1}$ assuming a $\Gamma = 1.7$ corrected for Galactic absorption.
\\
$^f$ Sources detected simultaneously in the soft and hard band. 
\\
$^g$ Sources detected in the soft band, but not in the hard band. 
\\
$^h$ Sources detected in the hard band, but not in the soft band.
\\
$^i$ $0.2 - 12\, keV$ band for sources detected at least in one of the standard detection bands
}
\end{flushleft}
\end{table*}

\subsubsection[]{$Log\, N(>S) - Log\, S$} \label{section4.3.2}
The source counts distribution  $(Log\, N(>S) - Log\, S)$  was obtained using our source catalog and the sky-coverage curves computed previously. The $Log\,  N(>S) - Log\, S$ represents the observed source counts $N(>S)$ as a function of the flux limits S of our survey, recovered from the sensitivity maps. We showed the form of the $Log\,N(>S) - Log\,S$ distributions using the integral source counts form $N(>S)$ as  the number of sources per unit of sky area  with measured flux higher than S:

 \begin{equation}
N\left( >{ S }_{ j } \right) =\sum _{ i=1 }^{ i=k }{ \frac { 1 }{ { \Omega  }_{ i } }  } 
\end{equation}

where $\Omega_i$  is the sky coverage (in $deg^2$) of the source $i$ in the bin, $S_j$ is the flux of the faintest element in the bin; the sum goes for the whole source list considering sources with flux $S_i > S_j$. Based on Poissonian statistics, the error bars are defined as $N(> S_j)/\sqrt{k}$  with $k$ as the total number of sources with $S_i  > S_j$. 

Nineteen out of our 301 sources, were detected in only one of the 13 observations, probably due to intrinsic variability. If they were detected during X-ray luminosity bursts, their  fluxes would not be representative of their average luminosity and could therefore bias the source counts distribution of our survey towards artificially high fluxes. Hence, we decided not to include these sources in our $(Log\, N - Log\, S)$  analysis, and we used only 282 X-ray objects that have been detected in at least two observations (see Table~\ref{table3}).

In Figure~\ref{logNlogS} we present our $Log\, N(>S) - Log\, S$ distribution for the hard and soft bands. We compared our source counts cumulative distribution  with  previous XMM-Newton and Chandra surveys, such as: \citet{Luo2017} with  the 7 Ms  Chandra deep field-south survey with a small coverage ($0.13\, deg^{2}$) and \citet{Civano2016} with the 4.6 Ms COSMOS-Legacy survey  ($2.2\, deg^2$) (Chandra). Then, we compared with \citet{Carrera2007} and \citet{Chen2018}  (XMM-Newton) for medium areas of $4.8$ and $5.3\, deg^{2}$, respectively. Additionally, for the hard band we compared with the 3.45 Ms XMM-Newton deep survey in the CDF-S \citep{Ranalli2013} with a sky coverage of $28.8\arcmin \times 28.8\arcmin$, comparable with our field of $29.5\arcmin \times 29.5\arcmin$. 

Our $Log\, N(>S) - Log\, S$  distributions are in good agreement with the results of the aforementioned works, except for the hard X-ray ``bump'' at $1 - 4\times 10^{-14}\, erg\, cm^{-2}\, s^{-1}$, which was also seen by \citet{Puccetti2009}. This deviation might be due to low counting statistics induced by our small sky coverage survey (cosmic variance) plus the effects of the difference in the cross-calibration for each survey and the spectral model used for the flux estimation.  A summary of our source counts cumulative distribution is presented in Table~\ref{table3} for the soft, hard, and full bands.

\begin{table}[hbt!]
\caption{Summary of $Log\, N(>S) - Log\, S$ distribution for Soft, Hard, and Full-band bands, respectively.}
\label{table3}
\scriptsize
\tiny
%\resizebox{\linewidth}{!}{%
\centering
\begin{tabular}{ccccccc}
\hline \hline 
Flux$^a$  & \textbf{$N(>S)$}$^b$  &\textbf{$N$}$^c$ & \textbf{$N(>S)$}$^b$  & \textbf{$N$}$^c$ & \textbf{$N(>S)$}$^b$ & \textbf{$N$}$^c$ \\
\textbf{ (S)} &  \textbf{Soft} & &  \textbf{Hard}  & & \textbf{Full-band}  & \\
\hline
$3.71\times 10^{-16}$ & $1107\pm$ 67 & \textbf{269} & - & - & - & - \\
$6.75\times 10^{-16}$ & $1015\pm$ 63 & 260 & - & - & - & - \\
$1.23E\times 10^{-15}$ & $845\pm$ 57 & 220 & $1753\pm$ 123 & \textbf{205} & - & - \\
$2.23\times 10^{-15}$ & $568\pm$ 49 & 132 & $1313\pm$ 94 & 197 & $802\pm$ 58 & 196 \\
$4.05\times 10^{-15}$ & $284\pm$ 34 & 68 & $852\pm$ 63 & 181 & $745\pm$ 54 & 187 \\
$7.36\times 10^{-15}$ & $140\pm$ 24 & 35 & $562\pm$ 49 & 130 & $626\pm$ 50 & 156 \\
$1.34\times 10^{-14}$ & $85\pm$ 19 & 21 & $280\pm$ 34 & 67 & $367\pm$ 38 & 91 \\
$2.43\times 10^{-14}$ & $21\pm$ 9 & 5 & $152\pm$ 25 & 36 & $185\pm$ 27 & 46 \\
$6.04\times 10^{-14}$ & $4\pm$ 4 & 1 & $24\pm$ 10 & 6 & $48\pm$ 15 & 10 \\
$1.10\times 10^{-13}$ & - & - & $4\pm$ 3 & 1 & $9\pm$ 5 & 3 \\ 
\hline \hline
\end{tabular}
%}
\begin{flushleft}
\tablecomments{\\ \footnotesize
$^a$ Flux limits in $erg\, s^{-1} cm^{-2}$.
\\
$^b$ Source counts per $deg^2$ by band with poissonian error.
\\
$^c$ Cumulative number of filtered sources (as described in the text) observed with fluxes corrected for galactic absorption higher than the flux limits.}
\end{flushleft}
\end{table}

\subsection{Hardness Ratio and obscured sources} \label{section4.2}

The hardness ratio $(HR)$ is a powerful indicator of the intrinsic spectrum of an X-ray source. The $HR$ value can also indicate the amount of obscuration by assuming a simple power-law model. The $HR$ is defined as follows: 

\begin{equation}
HR = \frac{H-S}{H+S}
\end{equation}

where S are the  soft band  count rates $(0.2 - 2\, keV)$ and H are the hard band count rates $(2 - 12\, keV)$. 

The source count rates used in our analysis are supplied by \texttt{edeteck$\_$stack} through the task \texttt{emldetect}. We considered the total counts from the PN camera in the 13 observations and the total cleaned exposure time corrected for  vignetting. In our analysis, we used a threshold limit of  $HR \ge -0.2$ to distinguish between  unobscured sources or type 1   \citep{Gilli2007} and obscured sources or type 2  \citep{Szokoly2004,Marchesi2016}. This threshold is also used in previous works,  e.g. \citet{Brusa2010},  who used multiwavelength  observations on the XMM-Newton survey of the COSMOS field.

\begin{figure}[!ht]
\includegraphics[scale=.39]{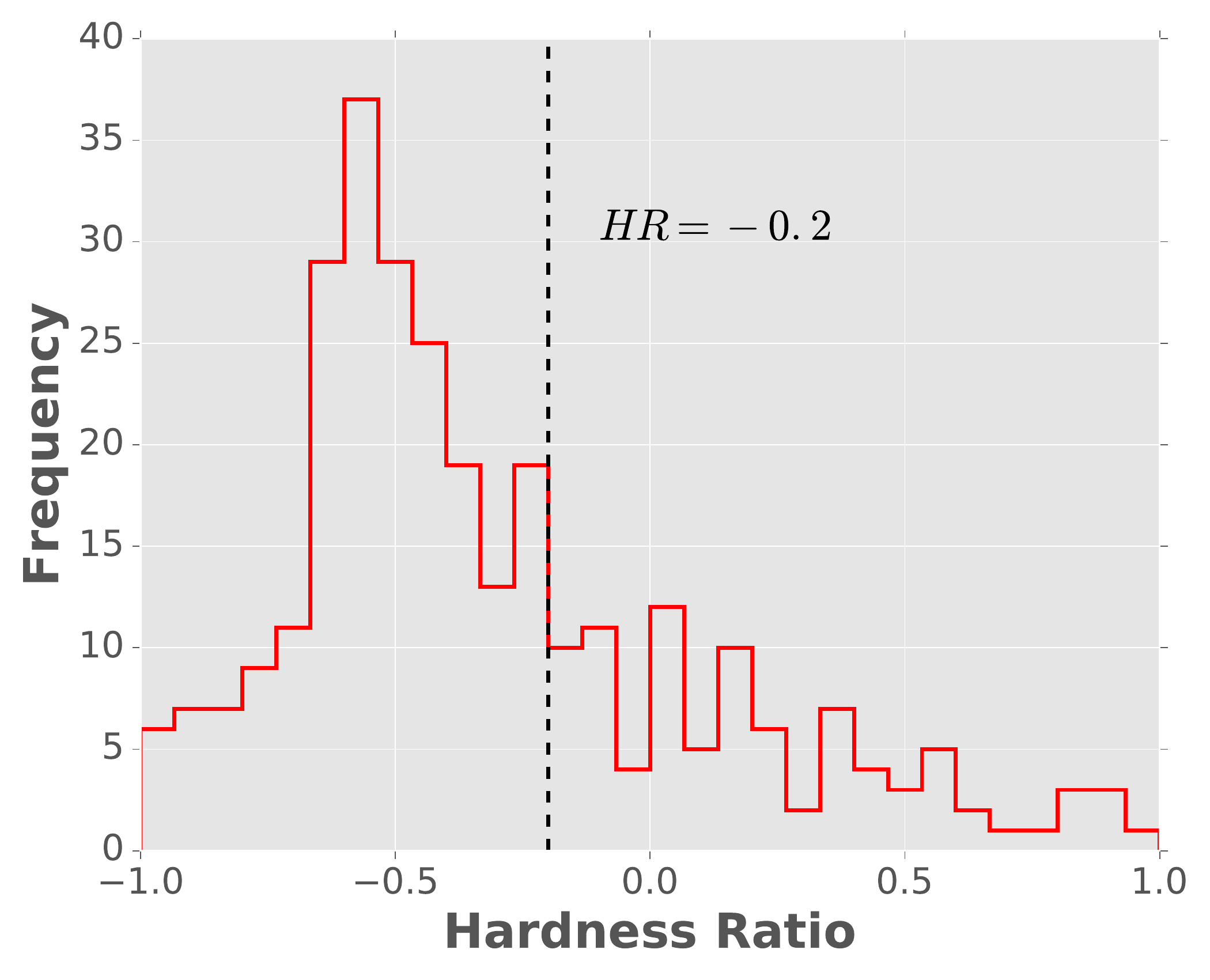} 
\caption{$HR$ distribution of our 301 sources, the black dashed line at $HR = -0.2$ separate between obscured (90) and unobscured sources (211).}
\label{fig6}
\end{figure}

In Figure~\ref{fig6}, we present the $HR$ distribution of our sample by using the mentioned $HR$ threshold. We found that $30\%$ (90) of our sources are obscured, whereas $70\%$ (211) are unobscured. The mean $HR$ of our sample is $HR = -0.31\pm 0.41$.

\begin{figure}[!ht]
\centering
\includegraphics[scale=.29]{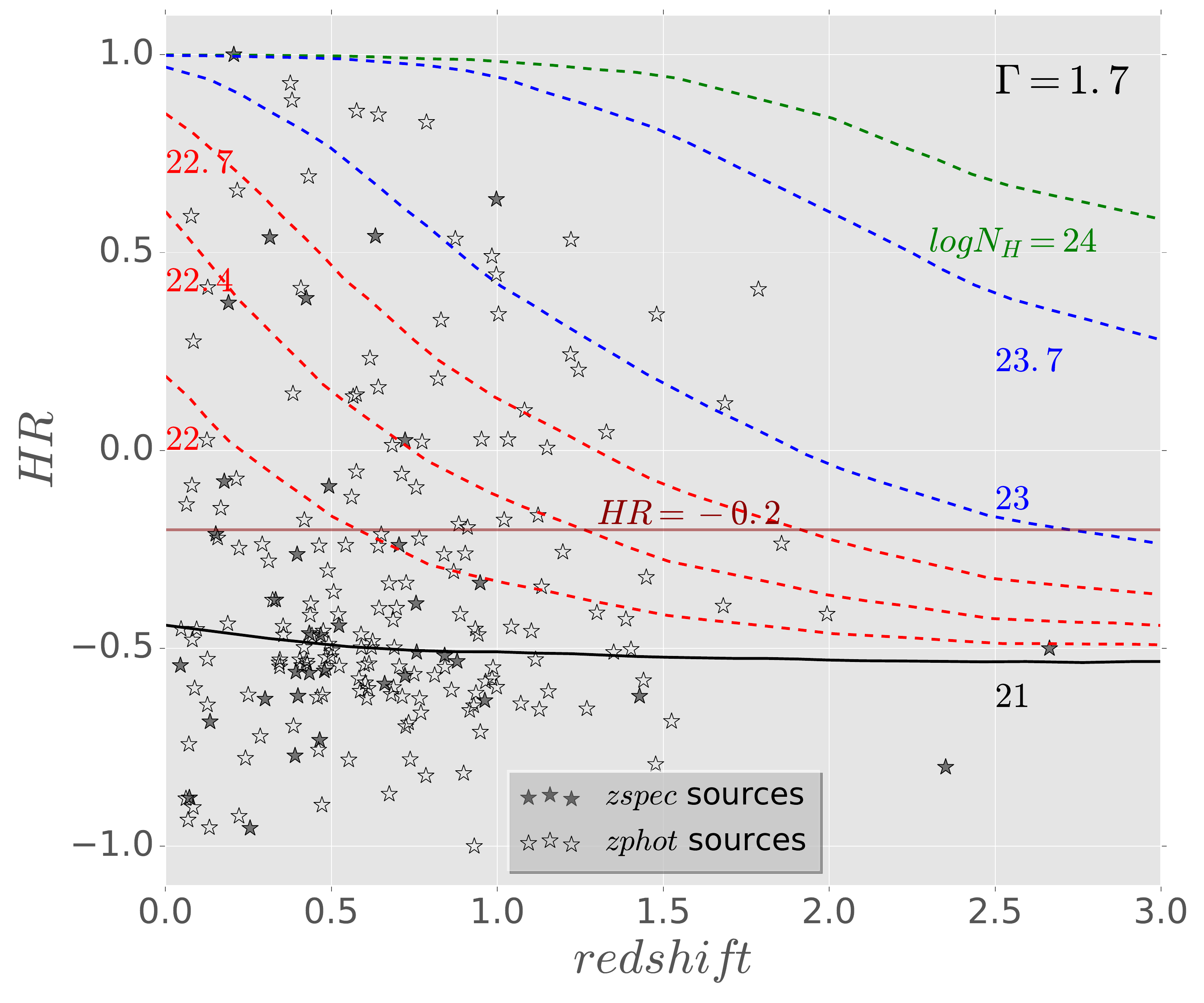} 
\caption{Hardness ratio vs redshift distribution of 211 sources, dotted lines mark different obscuration levels with $LogN_{H} = 24,\, 23.7,\, 23,\, 22.7,\, 22.4,\, 22$, and $21$, calculated assuming a spectral index of $\Gamma = 1.7$.}
\label{HR_z}
\end{figure}

Figure \ref{HR_z} shows the $HR$ vs $z$ distribution of our catalog. Following \citet{Elvis2012}, we included 7 curves  for different levels of obscuration ($logN_{H}=24,23.7,23,22.7,22.4,22,21$), assuming a constant spectral index of $\Gamma = 1.7$ and  adopting the PN response (QRF) corresponding to the cycle when these observations were taken. We observed that 62 out of 211 sources  ($29.4\%$) present obscuration with $logN_{H}  > 22$. 

\section{X-RAY AND OPTICAL/INFRARED RESULTS} \label{section5}

We calculated the X-ray luminosities ($Lx$) of our catalog from the observed flux at soft $(0.2 - 2)$, hard $(2 - 12)$, and full $(0.2 - 12)$  X-ray bands, assuming a $\Gamma = 1.7$ power-law spectrum corrected for Galactic absorption (see Section \ref{section4}). Moreover, following \citet{Xue2011} and \citet{Trouille2011}, we applied a K-correction with the equation:

\begin{equation}
Lx = Fx \times 4\pi D_{L}^{2} \times k_{correction}  
\end{equation}

where $k_{correction} = (1 + z)^{\Gamma - 2}$, $D_{L}$ is the luminosity distance and $Fx$ is our  X-ray flux. As noted before (Section \ref{section4.1}), we did not apply any further correction for intrinsic absorption in the luminosities reported here.

\subsection{Source type and AGN identification}\label{subsec5.1}

 We identified a subsample of AGN candidates from our X-ray catalog by using the criteria presented by \citet{Luo2017} updated from \citet{Xue2011} and used by \citet{Chen2018} in the XMM-SERVS survey. These three criteria are based on X-ray luminosity, optical/X-ray, and near-IR/X-ray flux ratios. When an X-ray source satisfies at least one of them, we classify it as an AGN candidate.   
 
\begin{enumerate}
\item An X-ray luminosity threshold $Lx > 3 \times 10^{42}\, erg\, s^{-1}$.  
%for high X-ray luminous sources.
\item An X-ray to optical flux ratio threshold of  $Log_{10}(Fx/Fr)> -1$. 
\item An X-ray to near-IR flux ratio threshold of  $Log_{10}(Fx/Fks)> -1.2$. 
\end{enumerate}

\begin{figure*}
\includegraphics[scale=0.3]{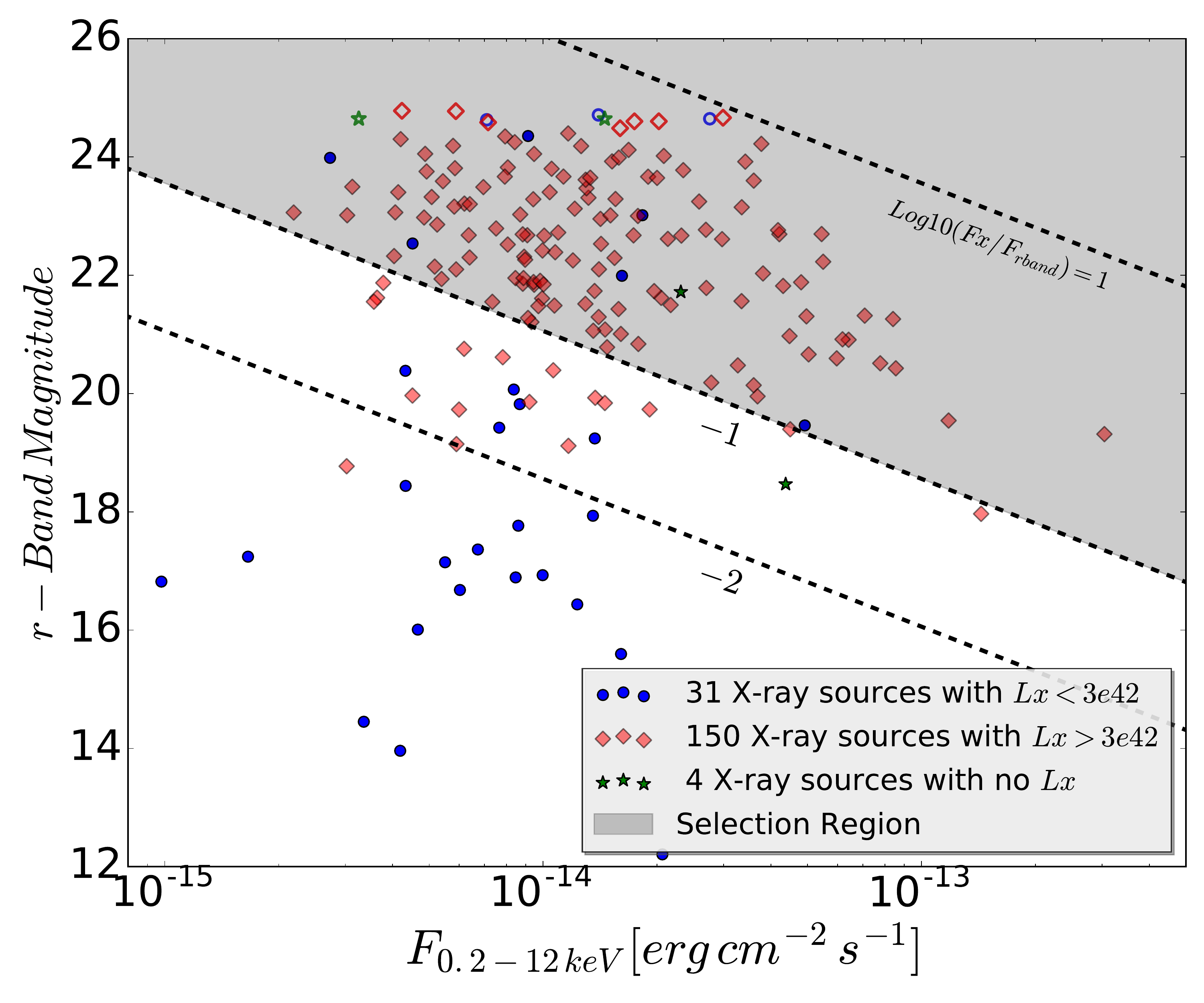} \includegraphics[scale=0.3]{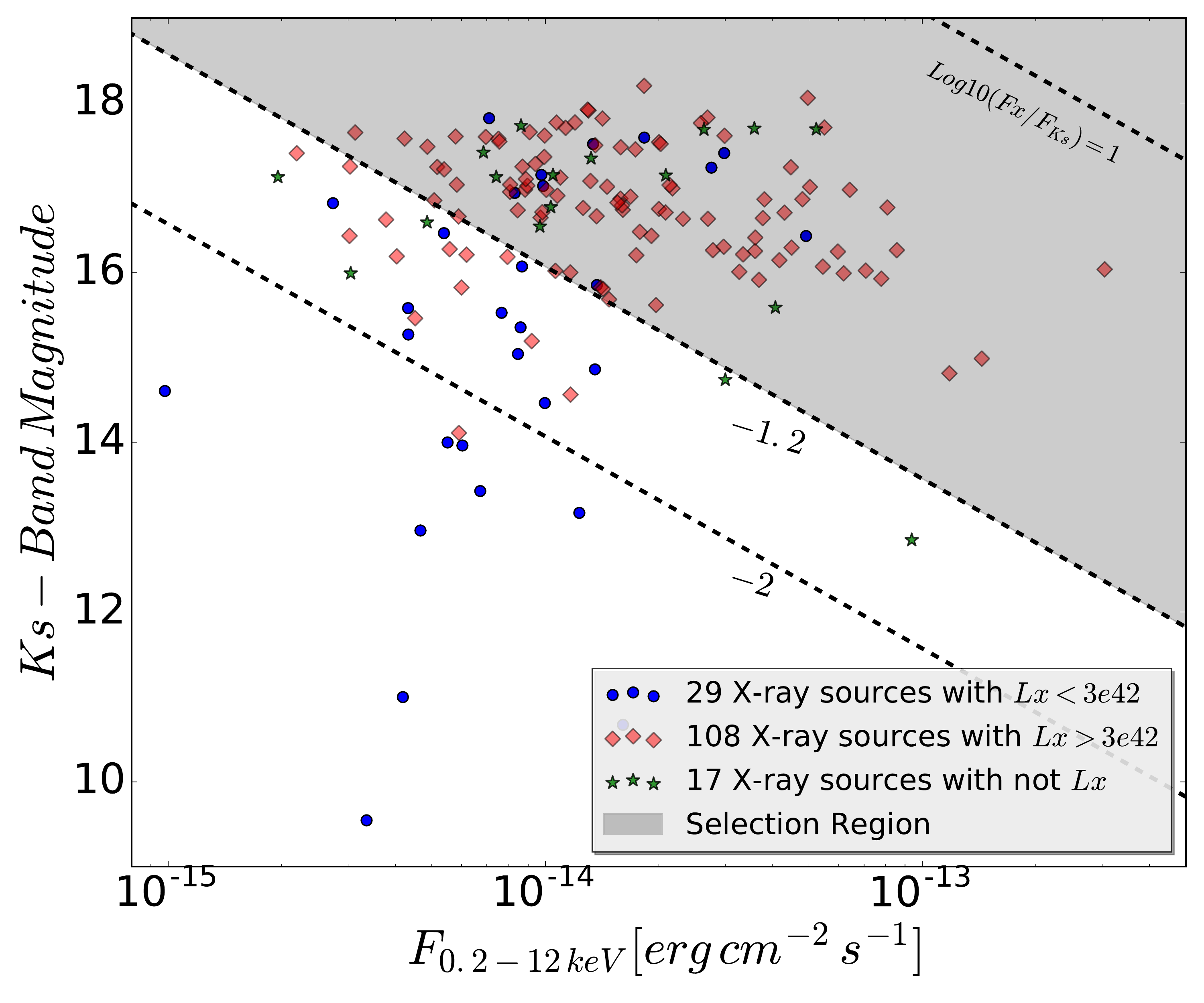} 

\caption{ Left: $F_{0.2 - 12\,keV} $ vs $r-band$  distribution for 185 sources from our X-ray catalog with optical counterparts. Filled (173) and unfilled symbols (12) mark those objects with magnitude $r'\le 24.4$ and $r'\le 24.8$ (upper limit), respectively. The black dotted lines represent the $Log_{10}(Fx/Fr)$ flux ratios at -2, -1, and 1, respectively, and the dark gray area mark the ``typical AGN selection region'' with $Log_{10}(Fx/Fr)\ge -1$.  Red diamonds, blue circles and green stars mark X-ray sources with $L_{0.2-12\,keV} < 3\times 10^{42}\, erg\, s^{-1}$, $L_{0.2-12\,keV} \ge 3\times 10^{42}\, erg\, s^{-1}$, and not estimation of $L_{0.2-12\,keV}$, respectively. Right: $F_{0.2 - 12\,keV} $ vs $Ks-band$  distribution for 154 X-ray sources with IR counterparts of our catalog, with the same symbols used in the left figure.}
\label{ratio}
\end{figure*}

According to the first criterion, we found 173 X-ray sources with $L_{0.2-12\,keV}>3\times 10^{42}\, erg\, s^{-1}$. For the second criterion, we found 147 objects with $Log_{10}(Fx/Fr)> -1$. Finally, for the third criterion we found a total of 117 sources with $Log_{10}(Fx/Fks) >-1.2$. To represent these results we show in Figure~\ref{ratio}  the $F_{0.2-12\,keV}$ vs $F_{r-band}$  distribution for 185 X-ray sources with magnitude  $r < 24.8$ (left-Figure) and the $F_{0.2-12\, keV}$ vs $Fks$ distribution for 154 sources of our catalog with infrared counterparts (right-Figure). In both plots the dark gray area represents the ``typical AGN  selection region'', while,  the red diamonds represent sources with $L_{0.2-12\,keV} \ge 3\times 10^{42}\, erg\, s^{-1}$ and blue circles are those with  $L_{0.2-12\,keV}<3\times 10^{42} \, erg\, s^{-1}$. Finally, by combining the three criteria we found that 204 $(\sim 84\%)$  of 244 sources are AGNs, of which  $50\%$ satisfy at least two criteria,  and $42\%$ satisfy all the three criteria. A redshift estimate is available for 184 out of the 204 AGNs identified here.

For a comparison, we downloaded \citet{Chen2018} catalog (hereafter \citetalias{Chen2018}) obtained with the XMM-SERVS survey from their webpage\footnote{\url{http://personal.psu.edu/wnb3/xmmservs/xmmservs.html}}. They used a catalog composed by observations of several instruments, such as: HSC-SSP $(r'<26.5)$, CFHTLS $(r'<24.8)$, and SDSS $(r'<22.5)$.  To compare our results with \citetalias{Chen2018} we used three flux ranges $10^{-15} - 10^{-13}\, erg\, cm^{-2}\, s^{-1}$, $10^{-15} - 10^{-14}\, erg\, cm^{-2}\, s^{-1}$  and $10^{-14}  - 10^{-13}\, erg\, cm^{-2}\, s^{-1}$, then we count the number of sources with $Log_{10}(Fx/Fr) > -1$ in both catalogs (considering the error propagation; see Table \ref{chen}).  Overall, our results are consistent with \citetalias{Chen2018}, i.e.  in the first range we found  $80^{+2.2}_{-6.1}\%$ of our sources with $Log_{10}(Fx/Fr) > -1$, while \citetalias{Chen2018} had $83^{+4.1}_{-5.1}\%$. On the other hand, for fainter sources ($10^{-15}  - 10^{-14}$ $erg\, cm^{-2}\, s^{-1}$) we detected discrepancies, mainly due to the magnitude limit used by \citetalias{Chen2018} of $r'<26.5$ with HSC, while we reached $r'<24.8$ for upper limit detection with GTC.

\begin{table*}
\caption{Comparison table of $Fx/Fr$ AGN criterion of  \citet{Chen2018} and our results. }
\label{chen}
\centering
%\resizebox{\linewidth}{!}{%
\begin{tabular}{@{}cccccc@{}}
\hline \hline
Flux range &  & Chen catalog & Chen catalog  & XMM175UNDF  & XMM175UNDF \\
$(erg\, cm^{-2}\, s^{-1})$& & & error $(+\sigma, -\sigma)$& $r'<24.8$& error $(+\sigma, -\sigma)$
\\ \hline
\multirow{3}{*}{$10^{-15} - 10^{-13}$} & Sources & 4887 & - & 180 & - \\
 & AGNs & 4057 & 4257, 3805 & 143 & 148, 133 \\
 & AGNs/Sources & 0.83 & 0.871, 0.779 & 0.8 & 0.822, 0.739 \\ \hline
\multirow{3}{*}{$10^{-15} -   10^{-14}$} & Sources & 2770 & - & 82 & - \\
 & AGNs  & 2248 & 2398, 2061 & 58 & 61, 50 \\
 & AGNs/Sources & 0.812 & 0.866, 0.744 & 0.7 & 0.744, 0.61 \\ \hline
\multirow{3}{*}{$10^{-14} -   10^{-13}$} & Sources & 2117 & - & 98 & - \\
 & AGNs & 1809 & 1859, 1744 & 86 & 87, 83 \\
 & AGNs/Sources & 0.855 & 0.878, 0.824 & 0.878 & 0.888, 0.847 \\ \hline \hline
\end{tabular}
%}
\end{table*}

\begin{figure*}
\centering
\includegraphics[scale=0.35]{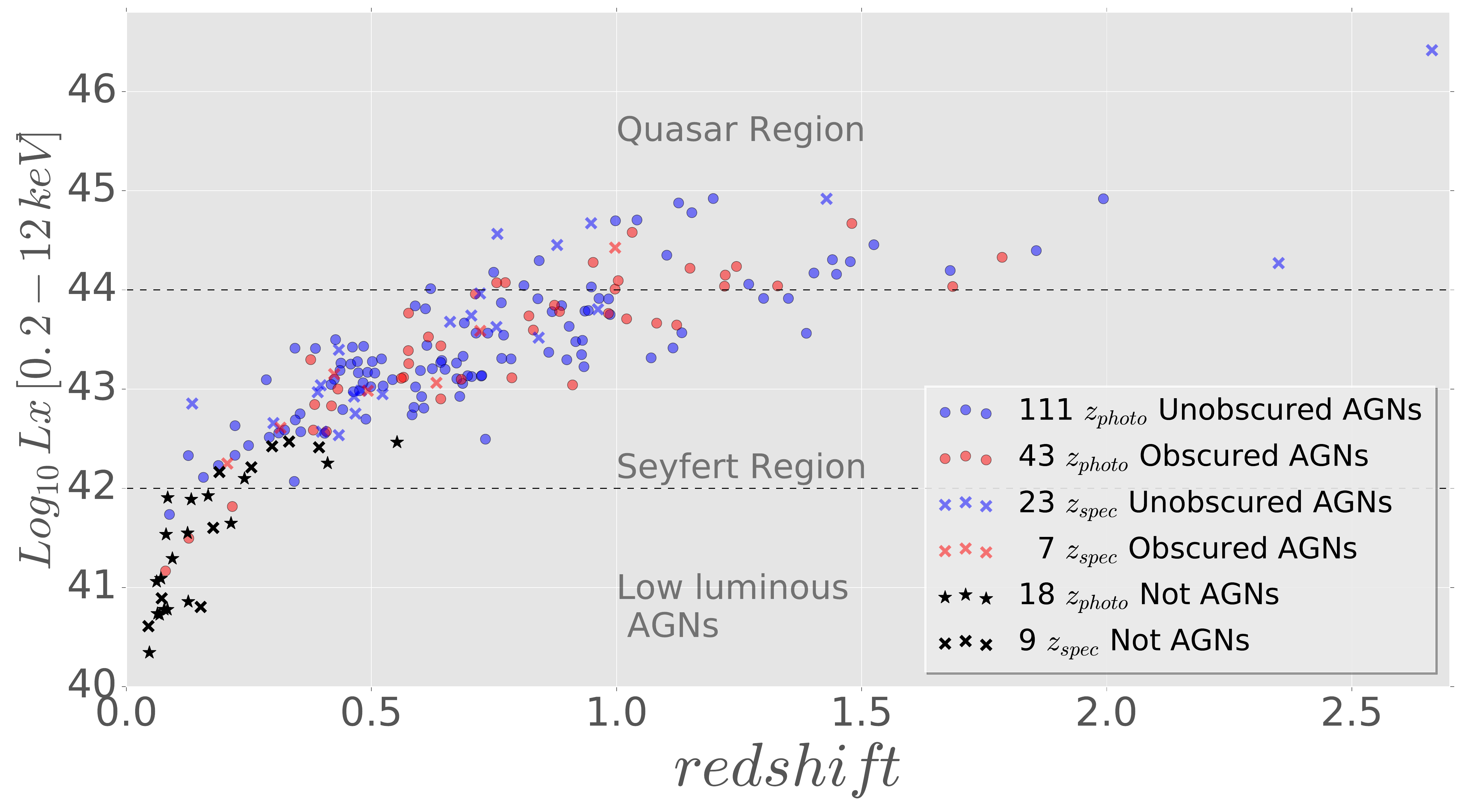} 
\caption{ $Redshift$ vs $Log_{10}\, Lx\, [0.2 - 12\, keV]$ distribution for 211 sources of our X-ray catalog. Red and blue symbols represent obscured (50)  and  unobscured AGNs (134), respectively, where circles/stars are for measurements with $z_{photo}$ and  crosses mark $z_{spec}$, respectively. Moreover, we sub-classified the sources as a function of their luminosity  as  Quasars $Lx > 10^{44}$, Seyfert $Lx = 10^{42} - 10^{44} $, and LLAGNs  $Lx < 10^{42}\, erg\,s^{-1}$.  The  black symbols represent 27 No-AGNs  X-ray candidates.}
\label{zlum}
\end{figure*}

In Figure~\ref{zlum}, we present the $redshift$ vs $Log_{10} (L_{0.2-12\, keV})$ distribution of our X-ray catalog, with spec-{\it z} sources (cross symbols, up to $z\sim2.7$)  and  photo-{\it z} sources (circles/stars symbols). We classified our targets into three broad luminosity groups: Low-Luminosity AGNs, with $Lx<10^{42}\,  \mathrm{erg\, s^{-1}}$, Seyferts, with $Lx=10^{42} - 10^{44}\,  \mathrm{erg\, s^{-1}}$, and Quasars, with $Lx>10^{44}\,  \mathrm{erg\, s^{-1}}$. In our sample, we count 139 Seyfert galaxies, 41 Quasars, and 4 LLAGNs (see Table~\ref{table4}). 

\begin{table}[hbt!]
\centering
\caption{AGNs classification resume for the whole XMM175UNDF catalog. The Quasar, Seyfert and LLAGN type is selected as defined in the text.}
\label{table4}
\scriptsize
%\resizebox{\linewidth}{!}{%
\begin{tabular}{lll}
\hline
\hline
Counterparts & Type & Number \\ \hline 
with  redshift  (211) & Quasar & 41 \\
 & Seyfert & 139 \\
 & LLAGN & 4 \\
 & No-AGNs selected$^a$ & 27 \\ \hline
No redshift (34) & Unclassified-AGN$^b$ & 20 \\
 & No-AGNs selected$^a$ & 14 \\ \hline
No Counterpart (57) & Unknown$^c$ & 57 \\ \hline 
 & Total & 301 \\ \hline \hline
AGNs & Criterion 1 & 173 \\
 & Criterion 2 & 147 \\
 & Criterion 3 & 117 \\ \hline \hline
\end{tabular}
%}
\begin{flushleft}
\tablecomments{\\ \footnotesize
$^a$ Sources which did not satisfy any AGN criterion %(e.g. normal galaxies). 
\\
$^b$ AGNs with no $redshift$ estimation. 
\\
$c$ Sources without optical counterpart.
}

\end{flushleft}
\end{table}

\begin{figure*}
\includegraphics[scale=0.38]{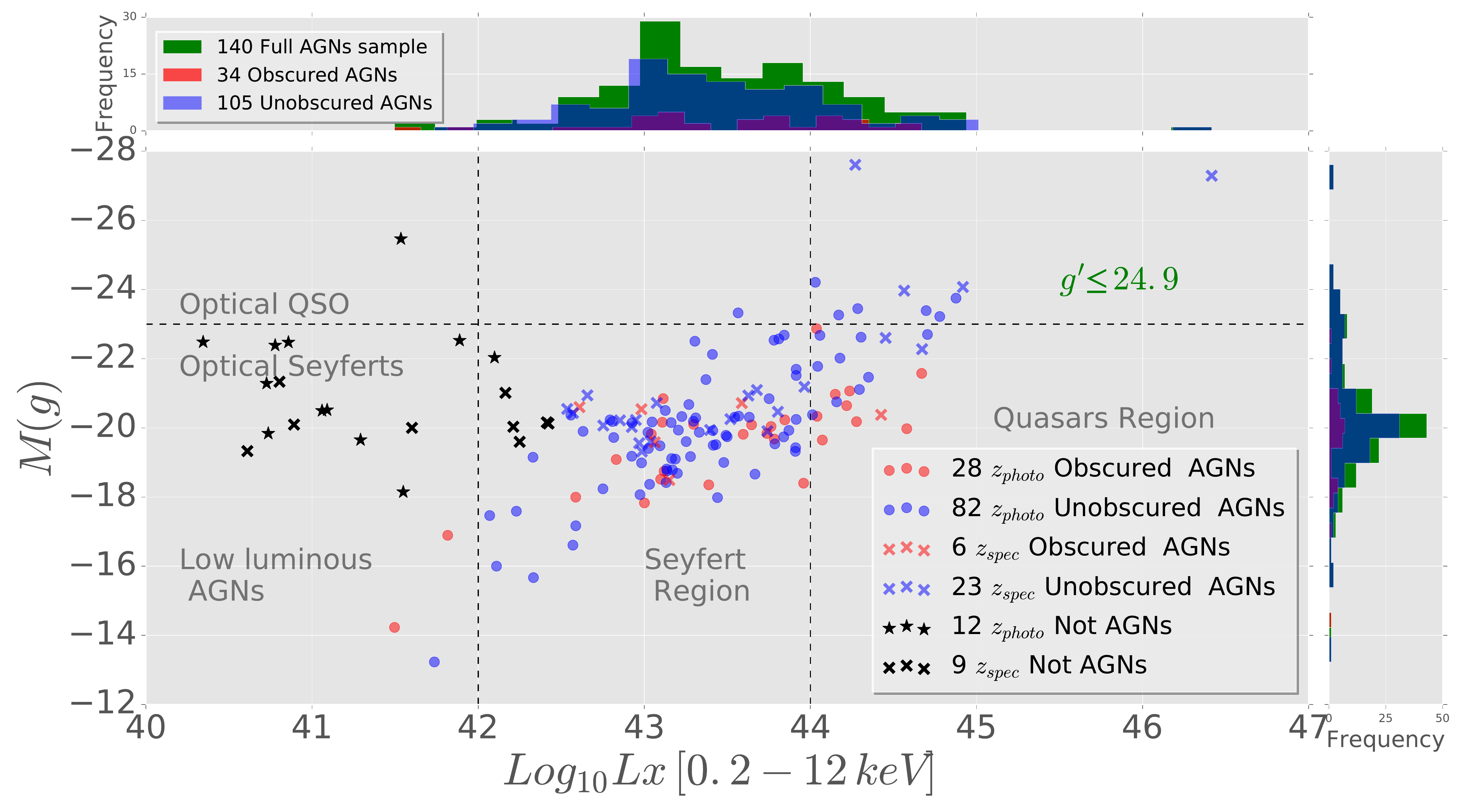} 

\caption{X-ray luminosity $L_{0.2-12\,keV}$ vs Absolute $g'$ band magnitude distribution of 160 X-ray sources with $g' \le 24.9$. Blue and red symbols represent unobscured (105) and obscured (34) AGNs, respectively, where circles/stars are for measurements with $z_{photo}$ and  crosses mark $z_{spec}$, respectively. Black symbols represent No-AGN candidates (21). The green, blue, and red histograms show  the dispersion of the full, unobscured, and obscured  AGN samples in function of $Lx$ and $M(g)$. The horizontal and vertical lines divide the chart in different regimes, the optical and X-ray transition between Seyferts, Quasars and/or QSOs.}
\label{diagnostic}
\end{figure*}

In Figure~\ref{diagnostic}, we show the distribution of the absolute $g'$ magnitude $M(g')$ vs $L_{0.2-12\, keV}$ of 131 source of our catalog with $g'<24.6$, we can observe a clear separation between AGNs (red and blue symbols) and non-active sources (dark symbols). Additionally,  we distinguish Seyfert galaxies and optical QSOs by using the equation $M_B = -21.5 + 5log(h) \approx - 23.3$  \citep{Schmidt1983,Schneider2006} 

\section{DISCUSSION} \label{discusion}

Following the analysis presented in the previous section, we selected a list of X-ray emitting AGNs.  From a subsample of 244 $(81\%)$ X-ray sources with optical/IR counterparts (301 detected in the XMM175UND-Field), we found a total of  204 AGNs, where $50\%$  of them  satisfied at least two of the three criteria outlined in Section \ref{subsec5.1}. This fraction increases to $90\%$ (219 AGNs) if we suppress  the  magnitude limit $r' < 24.8$ in the $Fx/Fr$ criterion.  This result is consistent with \citet{Chen2018} (by taking into account our differences in magnitude limits).

Possible causes for the lack of counterpart associations for the remaining (19$\%$) of our X-ray sources are: 1) the magnitude limit $(r' < 24.8)$ in our GTC catalog, 2)  the WISE cameras sensitivities and 3) possible spurious sources not yet removed from the catalog.
 
The absence of high z sources is likely due to the limited redshift of the bulk of our training set ($99\%$ up to $z \sim 1.5$). Therefore, to explore the possible high-z contents of our survey, it is necessary to proceed with a spectroscopic survey around the remaining 33 X-ray sources with Optical/IR counterparts without $z$ estimations.  Finally, the AGNs criteria selection used are highly reliable for luminous sources, but in some cases at low $Lx$ ($\sim 10^{42}\, erg\, s^{-1}$) we could be mis-classifying starburst galaxies as AGN candidates. Nevertheless, these results are in agreement  with the lack of bright Quasar and the  spatial distribution of  Seyfert galaxies population below $z\simeq 1$ \citep{fiore2003,paris2018}.

As a comparison, \citet{Marchesi2016} report a catalog of 4016 X-ray sources with a sky coverage survey 9 times larger than ours ($2.2\, deg^2$) in the 4.6 Ms Chandra COSMOS-Legacy Survey. They found a total of 1582, 717, 17, and $\sim 11$ sources at redshift ranges $z=1-2,\, 2-3,\, 4-5$, and $z > 5$, respectively.  These observations are the combination of 2 surveys, the 1.8 Ms C-COSMOS survey \citep{Elvis2009} and a 2.8 Ms Chandra observations \citep{Civano2016}. Their sensitivities of $1.9 \times 10^{-16},\, 7.3 \times 10^{-16}$ and $5.7 \times 10^{-16}\, erg\, s^{-1}\, cm^{-2}$ in the soft, hard, and full band, respectively, were obtained in a region of 0.5 $deg^2$ (two times wider and deeper than our survey) and 1.8 Ms (equivalent to our exposure). Considering these results and based on the assumption that the 11 sources at $z > 5$ are the faintest ones observed in the region with the highest sensitivity (0.5 $deg^2$) distributed homogeneously, we could expect  $\sim 4$ sources at $z>5$ in our survey by rescaling our sky coverage ($0.241\, deg^2$), exposure time (1.75 Ms) and sensitivities  with \citeauthor{Marchesi2016}.

On the other hand, \citet{Ranalli2013} with the 3.45 Ms XMM-Newton survey in the Chandra Deep Field South (similar sky coverage  $0.231\, deg^2$, twice nominal exposure time, and 2 times deeper in the hard band), reports a catalog of 339 sources at hard band with sensitivities of $6.6\times 10^{-16}\, erg\, cm^{-2}\, s^{-1}$ and 137 in super hard band $[5-10\, \mathrm{keV}]$ using a significance of $ML > 4.6$ (lower than our $ML > 6$). Since we found 212  hard and 82 super hard sources, these results are consistent  with \citeauthor{Ranalli2013}, considering lower exposure and higher $ML$.

The comparison of our source counts cumulative distribution with previous results showed an overall good agreement for different type of XMM-Newton and Chandra surveys in small and medium sky area coverage. Even so, there are small discrepancies in the $Log\, N(>S) - Log\, S$ in the  hard X-ray band.  These differences can be explained by the effect of low counting statistics (cosmic variance) due to the small sky coverage of our survey, plus possible effects of the difference in the cross-calibration and the spectral model used for each survey.

\section{SUMMARY AND CONCLUSIONS} \label{conclusiones}

In this paper, we present a deep XMM-Newton survey of the XMM175UND-Field, which consists of 13 observations centered on the same field of $30\arcmin \times 30\arcmin$ obtaining a total exposure time of 1.75 Ms (with cleaned PN of $1.372\, Ms$). An optical follow-up with the GTC telescope and a cross correlation analysis in optical and infrared bands allowed us to perform a multi-band study of our X-ray catalogue. A summary of our results is given below: 

\begin{enumerate}
 \item 	We computed the X-ray source detection using the new task \texttt{edetect\_stack} with the standard XMM-Newton bands $(0.2-0.5,\, 0.5-1,1-2,\, 2-4.5,\, 4.5-12\, keV)$ and significance threshold  of $p\simeq 2.5 \times 10^{-3}$  (equivalent to $\sim 3\sigma$). We obtained a reliable catalog of 301 X-ray point-like sources with flux limits of $4.03 \times 10^{-16}$, $1.3 \times 10^{-15}$ and $9.8 \times 10^{-16}\, erg\, s^{-1}\, cm^{-2}$  for the soft,  hard, and full band, respectively. Additionally, we did a detailed comparison analysis with the 4XMM catalogues of  \citet{Webb2020} and \citet{Traulsen2020}, resulting in a respective consistency of  $96\%$   and $94\%$  with both catalogs. 

 \item We used the LR technique to perform a cross-correlation analysis of our X-ray catalog with an optical catalog of $43,068$ objects produced by the OSIRIS instrument at GTC and an infrared-WISE public repository.  We were capable to detect optical/IR counterparts for 81$\%$ (244) of the whole XMM175UNDF catalog. %The remaining  57 sources are candidates for obscured and/or  high-redshift sources.
 
\item We computed our own photometric redshifts by using  \texttt{PhotoRApToR} with  a training set of 824 sources detected in our field (33 from our own GTC observations). About $99\%$ of our spec-$z$ catalog are contained in the range $z{spec} = 0 - 1.5$, thus we constrained our photo-{\it z} reliability up to $z = 1.5$. Then,  we achieved a $\sim 93\%$ of redshift completeness  for our  227 X-ray sources with optical counterparts. 

\item 	We calculated the $Log\,N(>S) - Log\,S$  distribution using the  sky coverage of our survey in a region of $0.241\ deg^{2}$. We found a general good agreement with previous XMM-Newton and Chandra surveys in small and medium areas. 

\item We obtained the $HR$ distribution of our source list and assuming a threshold of $HR \ge -0.2$, we found  that 30$\%$ (90) of the sources are obscured, of which $87.9\%$ have $logN_H  > 22$. We obtained a mean  and error $HR$ for the full catalog of $HR = -0.31\pm 0.41$. 

\item We used the criteria by \citet{Luo2017} to select  AGN candidates of our X-ray catalog based on their optical/IR and X-ray properties. We classified 204 objects as AGNs; of which 139 are Seyfert galaxies, 41 luminous Quasar, 4 LLAGNs, and 20 unclassified AGNs.
 
\end{enumerate}

\acknowledgments

YK acknowledges support from grant DGAPA-PAPIIT 106518 and from program DGAPA-PASP, and also acknowledges for the CONACyT  Project: A1-S-22784.
ALL and MEC acknowledge support from CONACyT grant CB-2016-286316.

%% To help institutions obtain information on the effectiveness of their 
%% telescopes the AAS Journals has created a group of keywords for telescope 
%% facilities.
%
%% Following the acknowledgments section, use the following syntax and the
%% \facility{} or \facilities{} macros to list the keywords of facilities used 
%% in the research for the paper.  Each keyword is check against the master 
%% list during copy editing.  Individual instruments can be provided in 
%% parentheses, after the keyword, but they are not verified.

\vspace{5mm}
\facilities{XMM-Newton, GTC(OSIRIS)}
%\facilities{HST(STIS), Swift(XRT and UVOT), AAVSO, CTIO:1.3m,
%CTIO:1.5m,CXO}

%% Similar to \facility{}, there is the optional \software command to allow 
%% authors a place to specify which programs were used during the creation of 
%% the manuscript. Authors should list each code and include either a
%% citation or url to the code inside ()s when available.

\software{The entire X-ray data set used in this article are available in the public XMM-Newton repositories, the ``XMM-Newton Science Archive'' on \url{http://nxsa.esac.esa.int/nxsa-web/\#home}.  The source code used for the cross correlation analysis Xcorr is available at \url{http://saada.u-strasbg.fr/docs/fxp/plugin/} \citep{pineau2011}.  While, PhotoRApToR is available at \url{http://dame.oacn.inaf.it/dame\_photoz.html}  \citep{Cavuoti2015} and the code used for our optical spectral reduction gtcmos version 1.4 is available at \url{https://www.inaoep.mx/~ydm/gtcmos/gtcmos.html} \citep{Gomez2016}.}
%\software{astropy \citep{2013A&A...558A..33A},  
%          Cloudy \citep{2013RMxAA..49..137F}, 
%          SExtractor \citep{1996A&AS..117..393B}
%          }

%% Appendix material should be preceded with a single \appendix command.
%% There should be a \section command for each appendix. Mark appendix
%% subsections with the same markup you use in the main body of the paper.

%% Each Appendix (indicated with \section) will be lettered A, B, C, etc.
%% The equation counter will reset when it encounters the \appendix
%% command and will number appendix equations (A1), (A2), etc. The
%% Figure and Table counter will not reset.
\newpage
\appendix
In this Appendix we present a more detailed explanation of some important steps during the preparation of this project, the idea is to simplify the understanding of:  1) the reliability of our results by presenting a deep comparison analysis of our X-ray catalog versus the newest XMM-Newton Survey Science Centre catalogs. 2) A precise description of the optical spectral analysis performed with optical sources detected in our own GTC observations over the XMM175UND-Field, which were used in the training set to estimate our photometric redshifts.  

\section{4XMM catalogs comparison} \label{ApendixA}
The source detection analysis on the XMM175UND-Field in Section \ref{section2} led to  a preliminary source list of 483 X-ray sources and a final catalog of 301 objects. 

During the completion of this work, two papers were published containing X-ray analyses of this same field: the newly obtained 4XMM-DR9 X-ray catalog for individual observations of \citet{Webb2020} (hereafter \citetalias{Webb2020}) and  the 4XMM-DR9s catalog for overlapping observations of \citet{Traulsen2020} (hereafter \citetalias{Traulsen2020}). 
Both catalogs were produced by XMM-Newton Survey Science Centre with $ML \ge 6$. First, \citeauthor{Webb2020} analyzed 14,041 individual observations from  February 2000  to  February  2019, finding 550,124 sources in an area coverage of $1152 \deg^{2}$. On the other hand,  \citeauthor{Traulsen2020} was constructed by using the same algorithm \texttt{edetect\_stack} employed in this work (see section \ref{section2}), they analyzed 1329 stacks with 6604 overlapping observations from  February 2000  to November 2018, finding 288,191 sources in an area of $300 \deg^{2}$. Both catalogs contain the same 13 pointings used in our analysis, with the addition of 10 extra PN small-window calibration observations (each of $\sim 30ks$) pointed on the blazar 1ES 1553+113 that our survey does not include. 

We present here a detailed comparison of our results with these two surveys. By using a cross correlation radius of $10\arcsec$ \citep[similar to][]{Chen2018} based on the positional accuracy, pointing uncertainties and PSF size of XMM-Newton observations; we exhaustively compare: 1) the maximum likelihood distribution of the X-ray sources, 2) the number of elements in each source list, their position and the possible reasons of discrepancy, 3) the flux distribution of sources in the three surveys. 

It is important to mention that the field studied in this paper is a very complex field due to the presence of a very bright source that induces bright spikes at the center of the images. 
This could easily result in a high number of spurious detections that can fluctuate between two catalogs. For this reason, in order to have a more reliable comparison between our catalog and the above catalogs, we will not include the sources which lie in the masking region used during our analysis and sources flagged as spurious. As a result of this choice our survey contains less sources (301) compared to both 4XMM catalogs but with a very similar number of detections in our preliminary source list: 483 objects (see section \ref{section2.3}), to be compared with 477 (\citetalias{Traulsen2020}) and 478 (\citetalias{Webb2020}). 
%As a result: \citetalias{Traulsen2020} reports 477 X-ray objects, while \citetalias{Webb2020} reports 478. 

\begin{figure}[!ht]
\centering
\includegraphics[scale=.24]{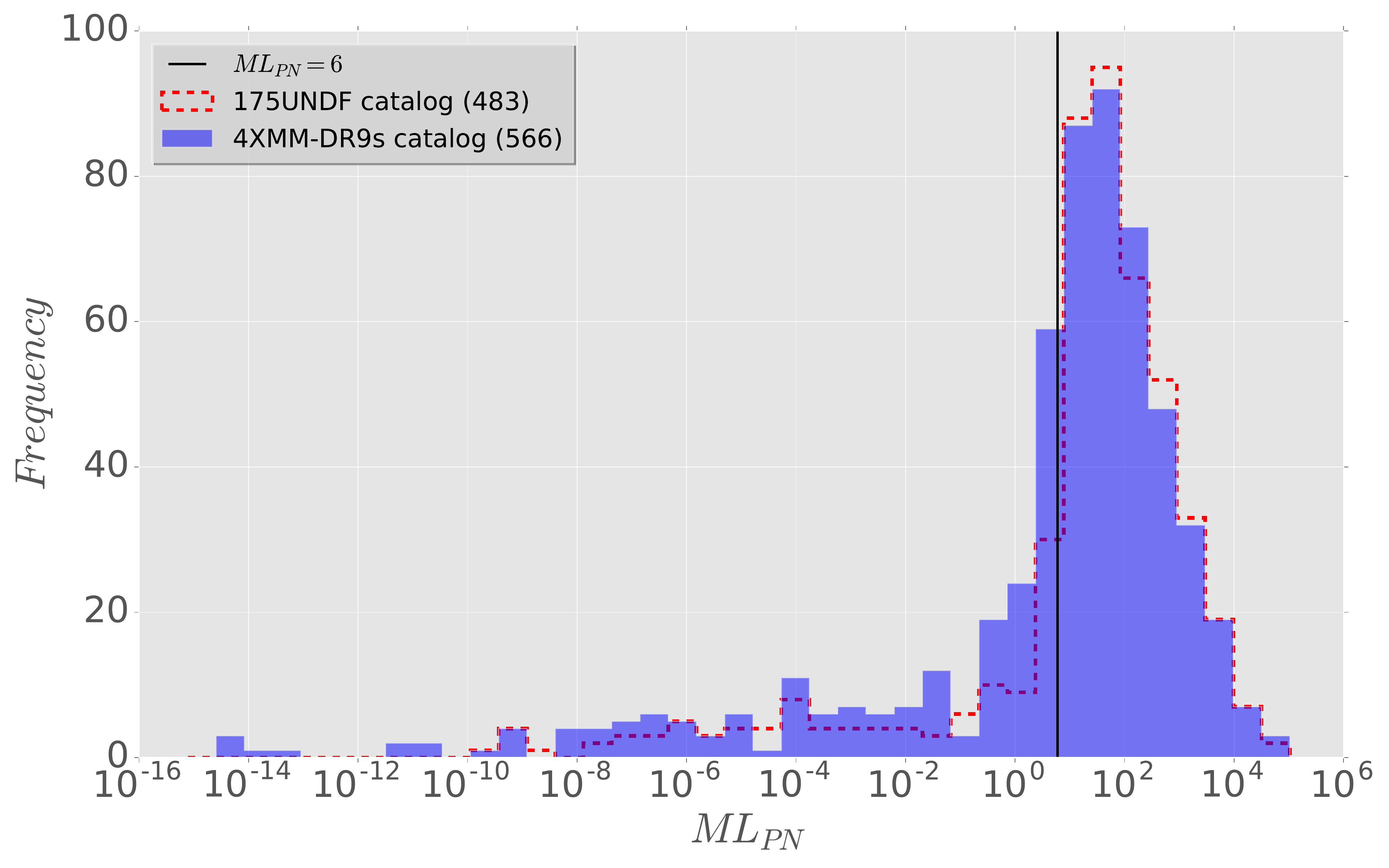} 
\caption{
Comparison of the detection significance in our preliminary source list with 483 sources (red) vs \citetalias{Traulsen2020} 566 objects (blue) of which 89 are flagged as spurious. We used the XMM-Newton EPIC/pn maximum likelihood distribution for both samples, the black vertical line refers when $ML_{PN} = 6$.}
\label{ap1}
\end{figure}

In Figure \ref{ap1}, we compared the detection significance for our source list and \citetalias{Traulsen2020}. At $ML>6$ we observe the same behavior for both samples, while at lower significance $(ML <6)$ the amount of possible spurious sources increase, which could modify the distribution observed in both catalogs. \footnote{Sources with $ML$ higher or equal to 6 in the final stacking or a at least in one observation in any band are select as X-ray source, i.e. there are X-ray sources selected which have $ML<6$ in the final stacking, but with a likelihood $\ge 6$ in at least one band in one or two individual observations \citep[standard selection technique of \texttt{edetect$\_$stack} see][for more details,]{Traulsen2019}.}

\begin{figure}[!ht]
\centering
\includegraphics[scale=.3]{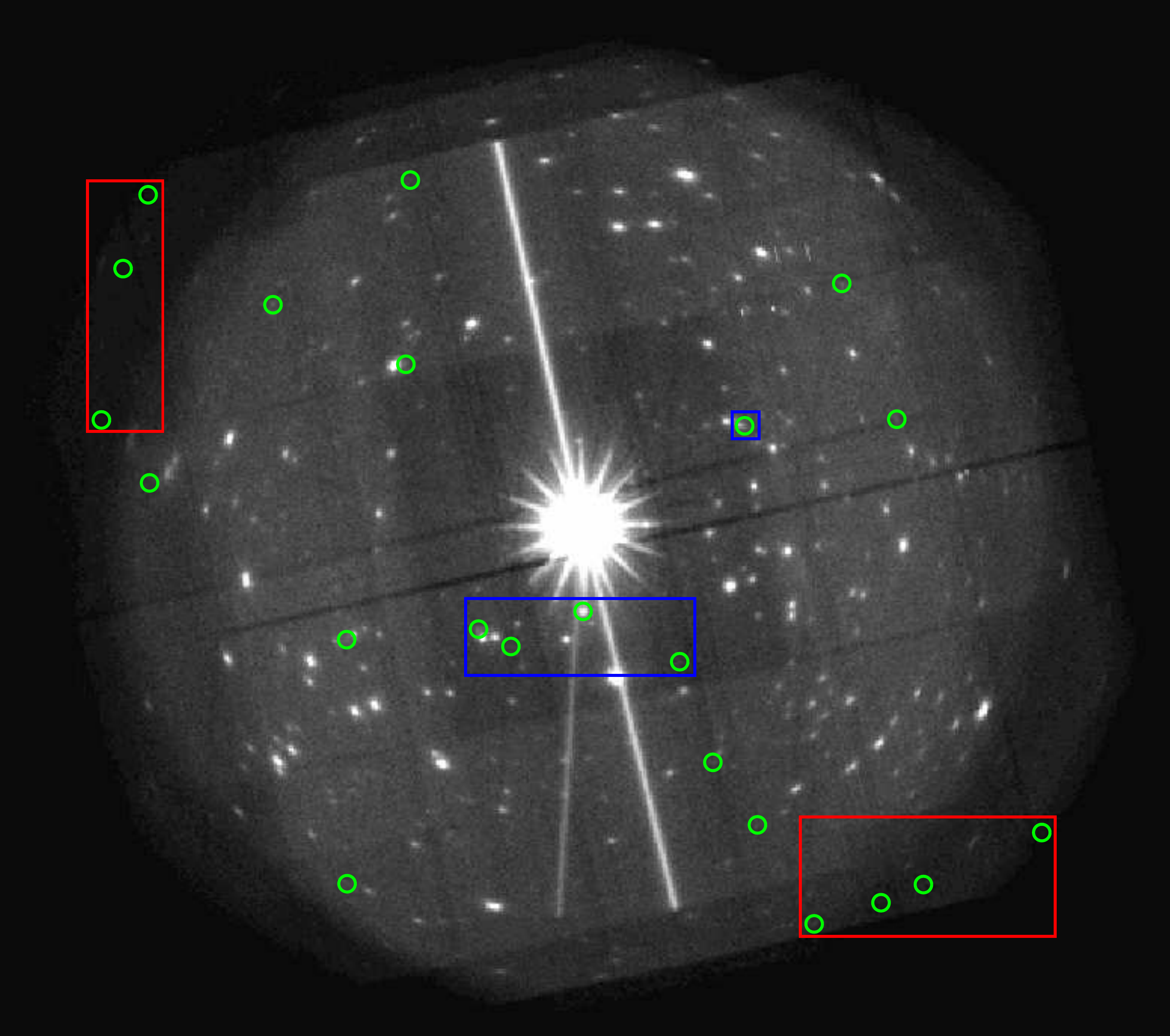} 
\caption{Mosaic image of all observations of the survey at full X-ray band $0.2 - 12\, \mathrm{keV}$. Green circles mark the 22 non-common sources detected in \citetalias{Traulsen2020} but not in our catalog. The red rectangles mark the position of 7 objects on the borders of the field, the blue rectangles mark the position of the 5 closest sources to the center of the field.}
\label{ap2}
\end{figure}

\begin{figure}
\centering
\includegraphics[scale=.24]{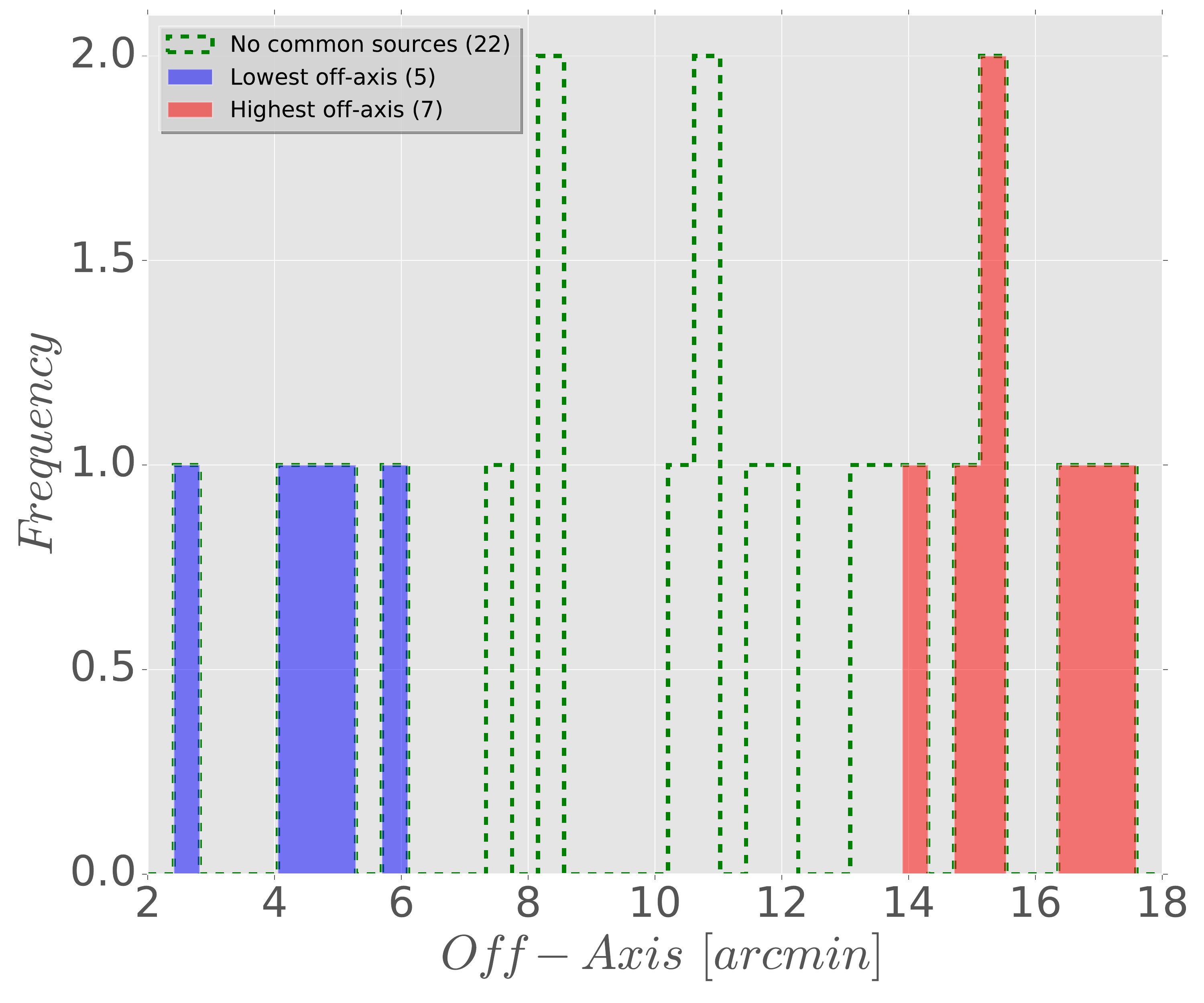}
\includegraphics[scale=.24]{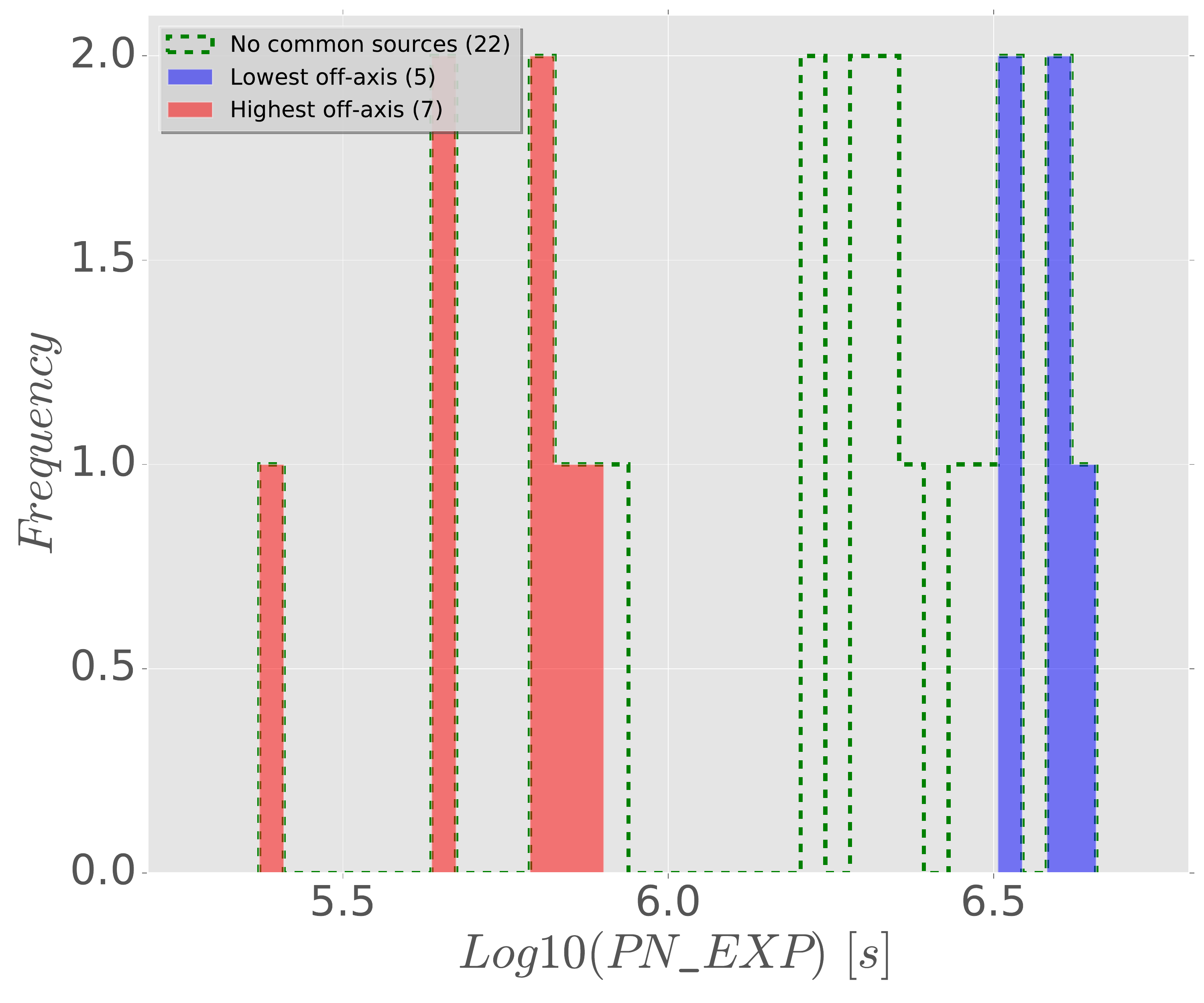} 
\caption{Distribution of 22 sources which are present in  \citetalias{Traulsen2020} but not in our catalog. Left: Off-axis distribution, Right: PN exposure distribution. In both figures the blue and red histograms refer to the lowest ($<6\arcmin$) and highest ($>14\arcmin$) off-axis angles.}
\label{ap3}
\end{figure}

Then, following  \citet{Chen2018},  we used a circular region of $10\arcsec$ to cross correlate both 4XMM catalogs with our preliminary source list (483) and final catalog (301). For \citetalias{Traulsen2020} (\citetalias{Webb2020}) we found 421 (406) and 288 (284) common X-ray sources with our preliminary source list and final catalog, respectively. 
In Figure \ref{ap2}, we show the spatial distribution of the non-common sample (22, green circles) along our XMM175UND-Field; the 7 sources with highest off-axis angle $>14\arcmin$ are highlighted  with red rectangles. These sources are also the ones with the lowest exposure, as expected $<740\, ks$, (see Figure  \ref{ap3}-left). These objects could be spurious sources in the borders of the detectors. The blue rectangles show the 5 highest exposure sources and the lowest off-axis angle $<6\arcmin$ (see Figure \ref{ap3}-right), three of those objects are likely faint sources detected thanks to  the combination of the stacking observations. There are likely some spurious objects $(\sim 4)$ detected in the wings of the PSF of a bright source. One exception is the source closest to the bright blazar, this is a clear X-ray source, which have optical and IR counterparts; since  this source lied in the region masked out during our source detection process, we did not include it in our final source list. 

\begin{figure}[!ht]
\centering
\includegraphics[scale=.24]{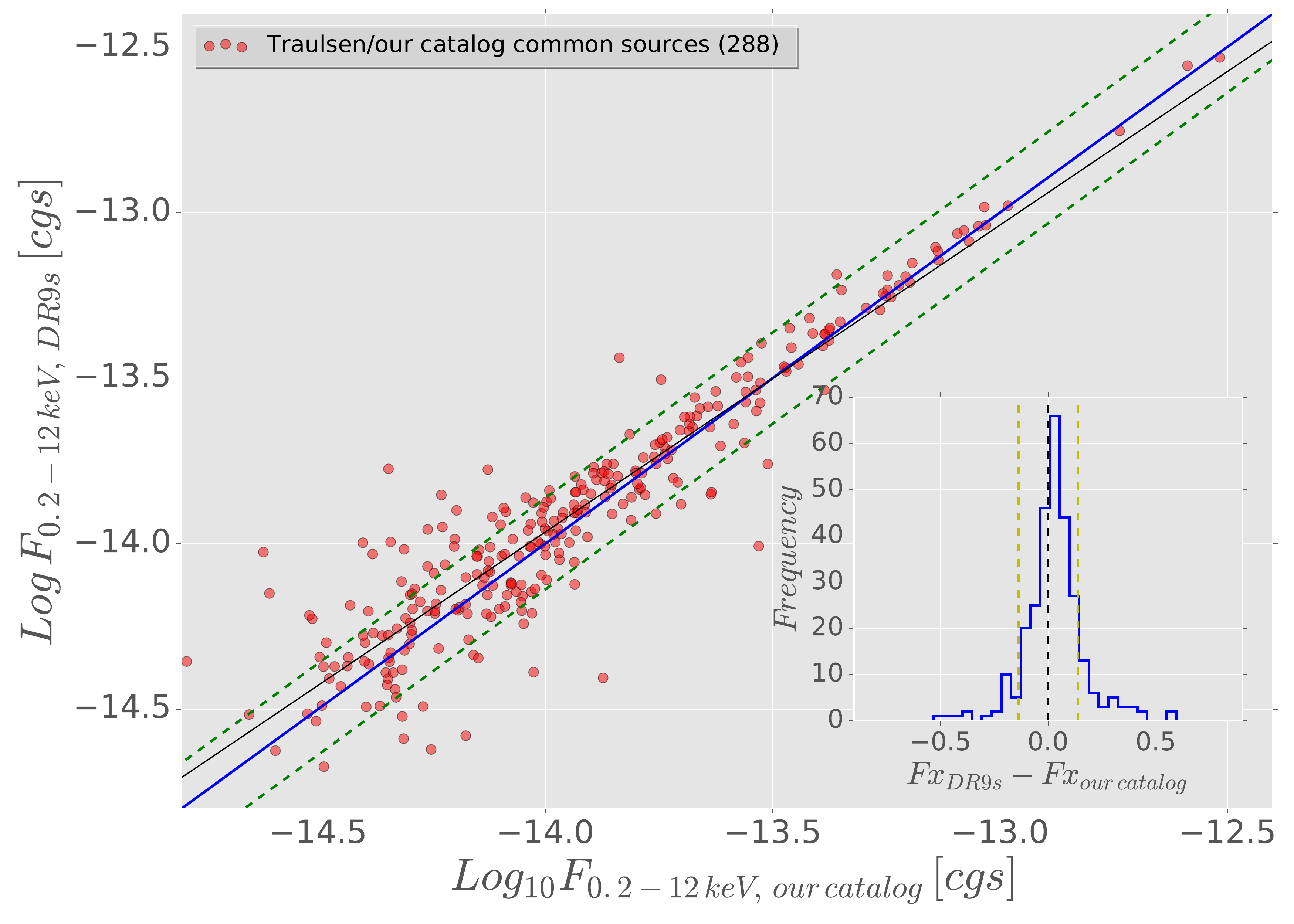} 
\includegraphics[scale=.24]{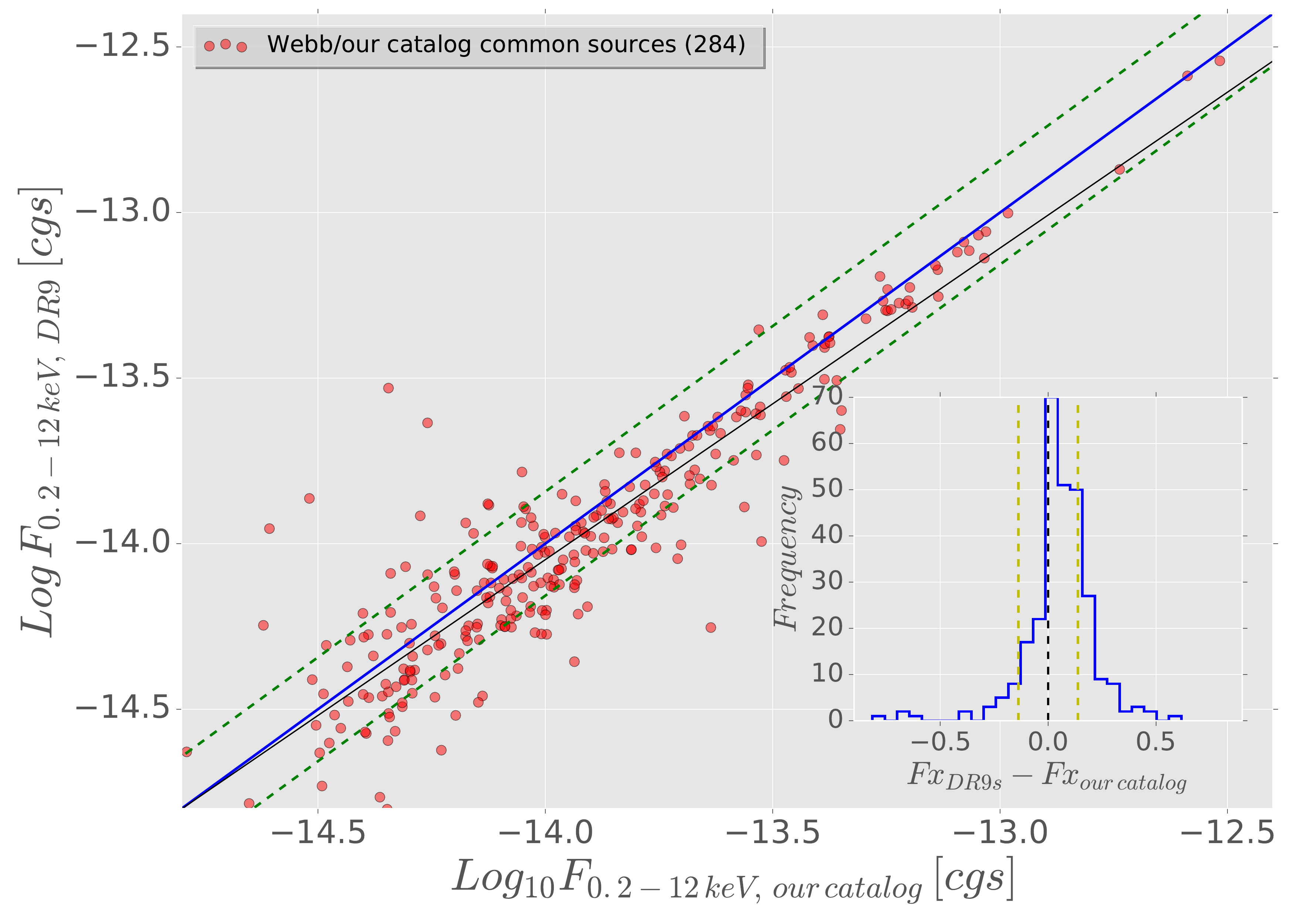} 

\caption{$0.2-12\, keV$ flux distribution for both 4XMM common sources catalogs with our final catalog. Left: DR9s vs our catalog flux distribution for 288 sources. Right: DR9 vs our catalog flux distribution for 284 sources. The blue line mark $F_{our catalog} = F_{DR9s}$ relation, the green dotted lines refers the $\sigma$ confidence locus. Right in the bottom we include the residual distribution for each plot.}
\label{ap4}
\end{figure}

Then, we perform a comparison of the X-ray fluxes obtained by the three catalogues. Figure \ref{ap4}  shows the EPIC-$Flux_{0.2 - 12\, keV}$   distribution of our common sources with \citetalias{Traulsen2020} (Left) and \citetalias{Webb2020} (Right). We found a good consistency with both 4XMM catalogs flux estimates (mainly with \citetalias{Traulsen2020}), such as: the standard deviation for both distributions are $\sigma_{DR9s}=0.06$  and $\sigma_{DR9}=0.07$, respectively. The best linear fit for the common source fluxes are expressed by the equations:

\begin{equation}
log10(y_{DR9s})= 0.4log10(x) - 0.99 \quad and \quad log10(y_{DR9})=  0.41log(x) - 0.87
\end{equation}

\begin{figure}[!ht]
\centering
\includegraphics[scale=.24]{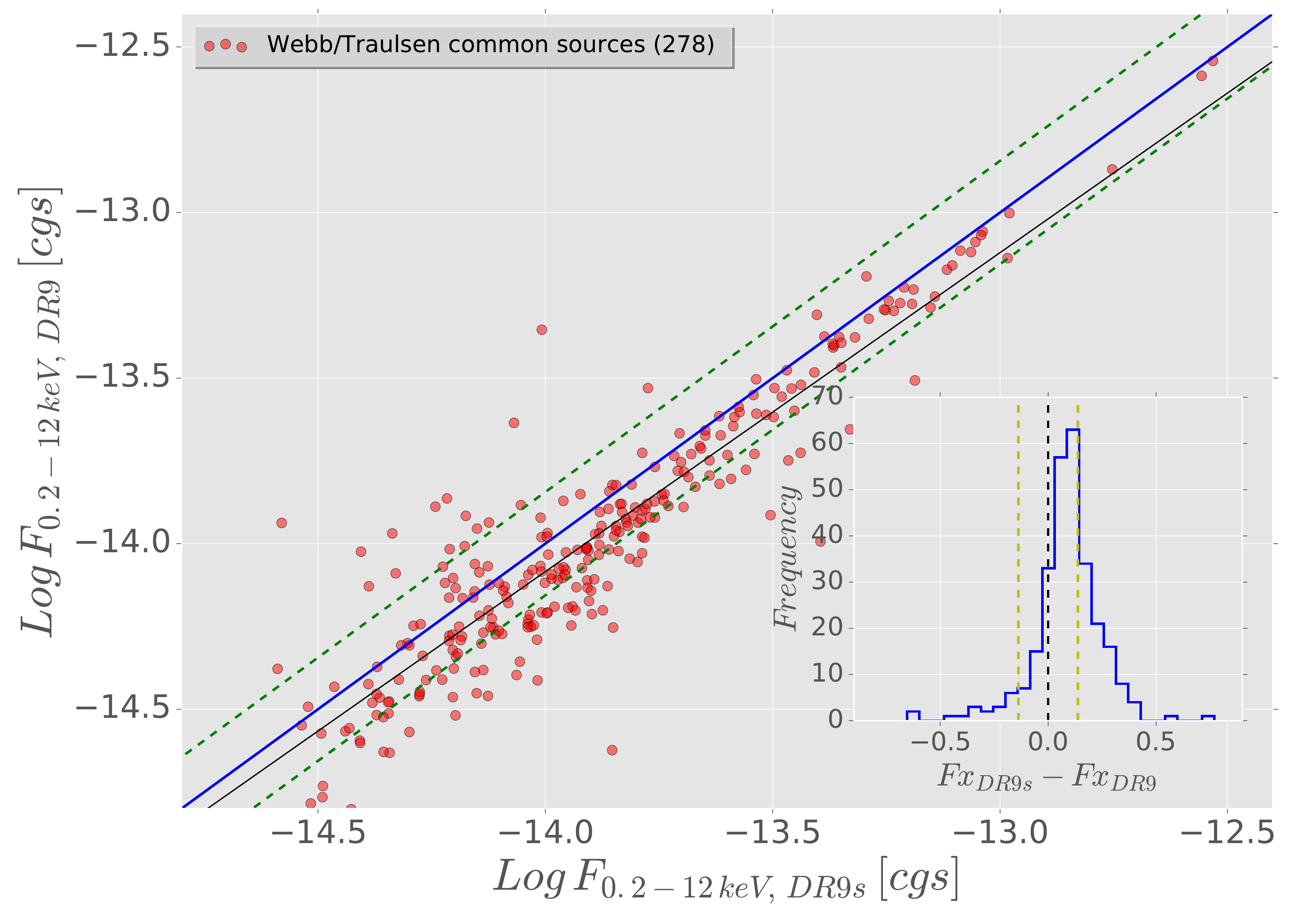} 
\caption{\citetalias{Traulsen2020} vs \citetalias{Webb2020} $0.2-12\, keV$ flux  distribution for 278 common sources in both 4XMM and our final catalog. The other elements are similar to Figure \ref{ap4}.}
\label{ap5}
\end{figure}

Due to the intrinsic variability of AGNs, \texttt{edetect\_stack} used in this work  and by \citetalias{Traulsen2020} can reduce the probability to underestimate or overestimate the real fluxes by computing the average flux for the whole overlapping observation for each source. Then, we computed the flux distributions of both 4XMM catalogs; in Figure \ref{ap5} we can see how \citetalias{Webb2020} obtains systematically lower fluxes than \citetalias{Traulsen2020}, which might be pointing to a real effect of a systematic  underestimation of fluxes in \citetalias{Webb2020}.  Then, the best linear fit for both 4XMM catalogs are:

\begin{equation}
log10(y) = 0.42log10(x) - 0.6
\end{equation}

After this analysis, we can conclude that our results are in satisfactory agreement with both 4XMM catalogs. As a summary, in Figure \ref{ap1} we found a comparable $ML$ distribution for our source list and \citetalias{Traulsen2020}. We found only 22 non-common sources with our catalog (most of them explained above, see Figure \ref{ap2}). Finally, we obtained a solid flux distribution correlation for both 4XMM catalogs (close to 1:1 for \citetalias{Traulsen2020}) with $\sigma_{4XMM-DR9s}=0.06$ and $\sigma_{4XMM-DR9}=0.07$ for \citetalias{Traulsen2020} and \citetalias{Webb2020}, respectively (see Figure \ref{ap4}). 

\section{GTC spectral analysis} 
\label{gtcmos}

Our spectroscopic observations were carried out with the MOS configuration of the instrument OSIRIS with GTC and consist of 5 observational blocks with 33 slits per block and $\sim6-7$ stars as fiducial points for astrometry and $1-2$ for sky spectral subtraction.

Every block was observed in three runs of 15 minutes (45 min. total exposure), reaching enough sensitivity to allow detection of emission and/or absorption lines in 33 out of $\sim 100$ sources and estimate their spectroscopic redshift. The observations were performed with the R1000R grism, centered at 7430 \AA\, covering the range from 5100 to 10000 \AA\, at a resolution of 2.62 \AA/pixel.

\begin{figure*}
\centering
\includegraphics[scale=.34]{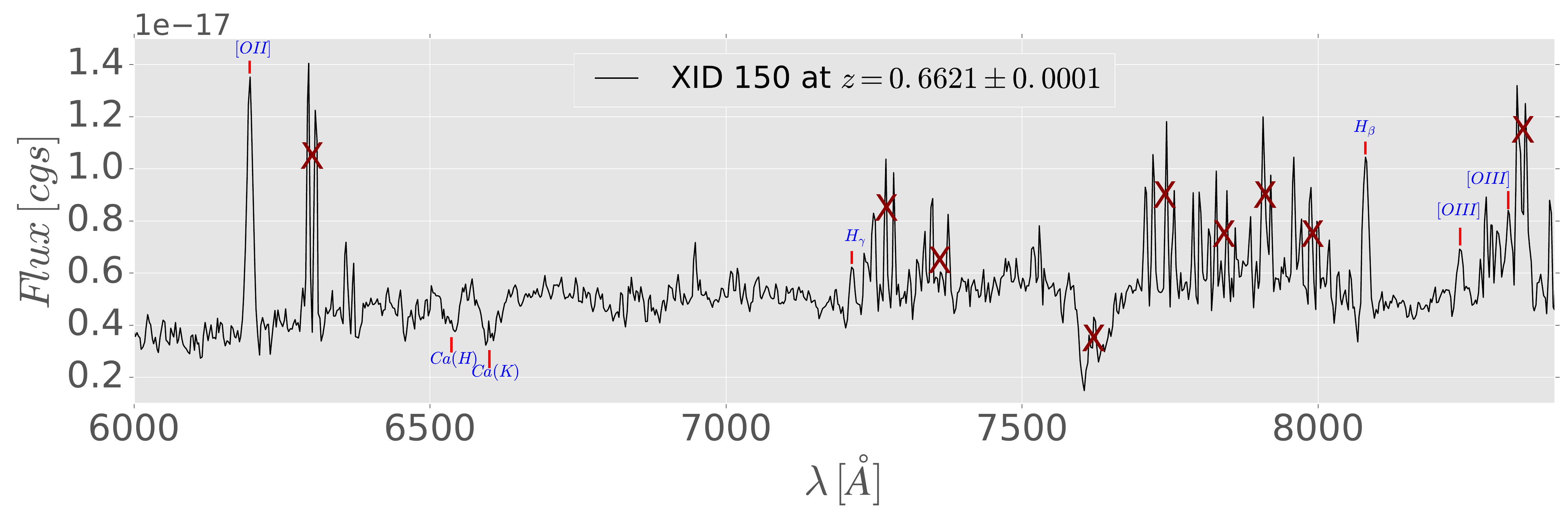}
\includegraphics[scale=.34]{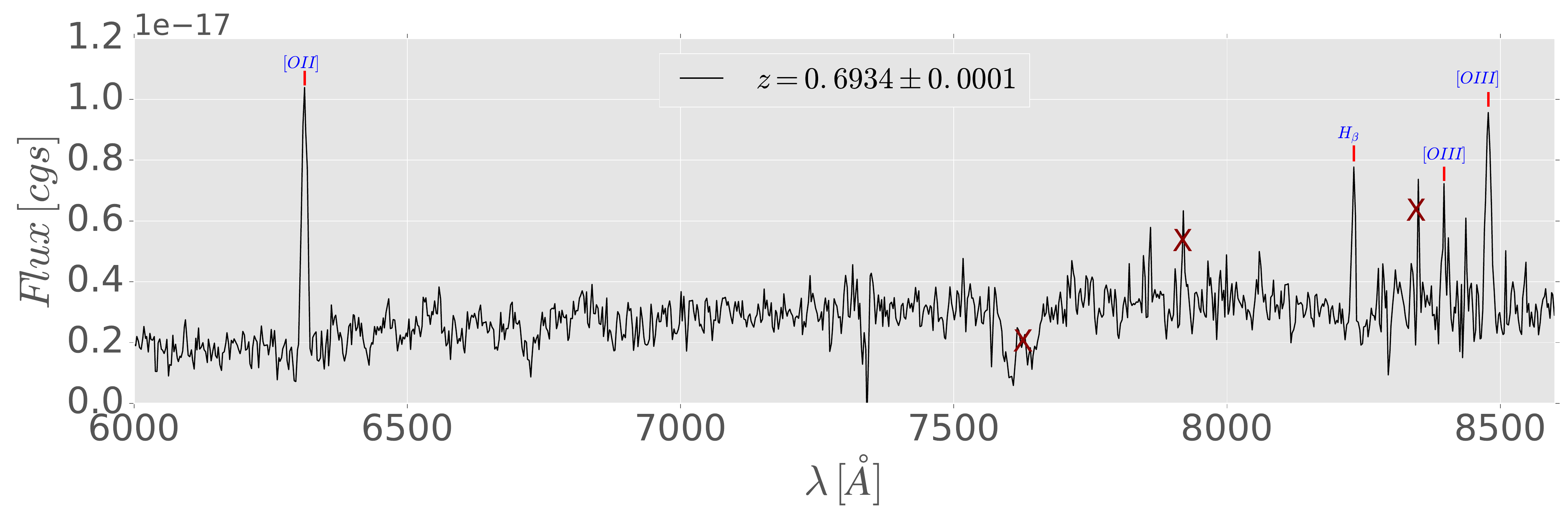}
\includegraphics[scale=.34]{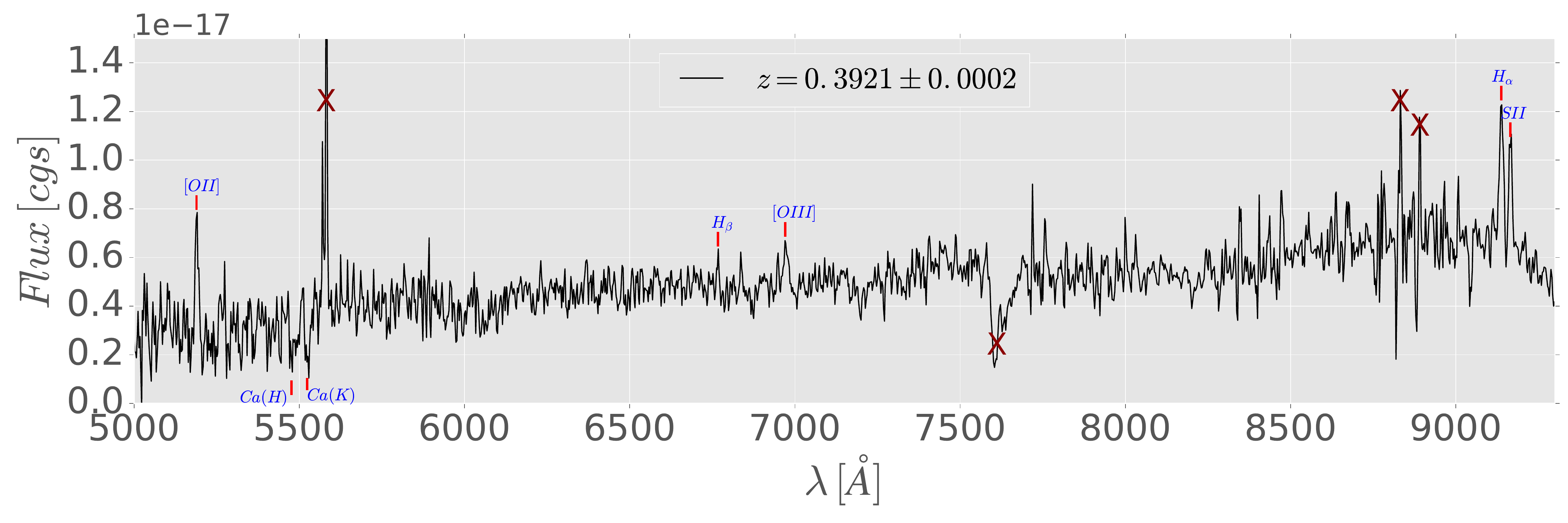}
\includegraphics[scale=.34]{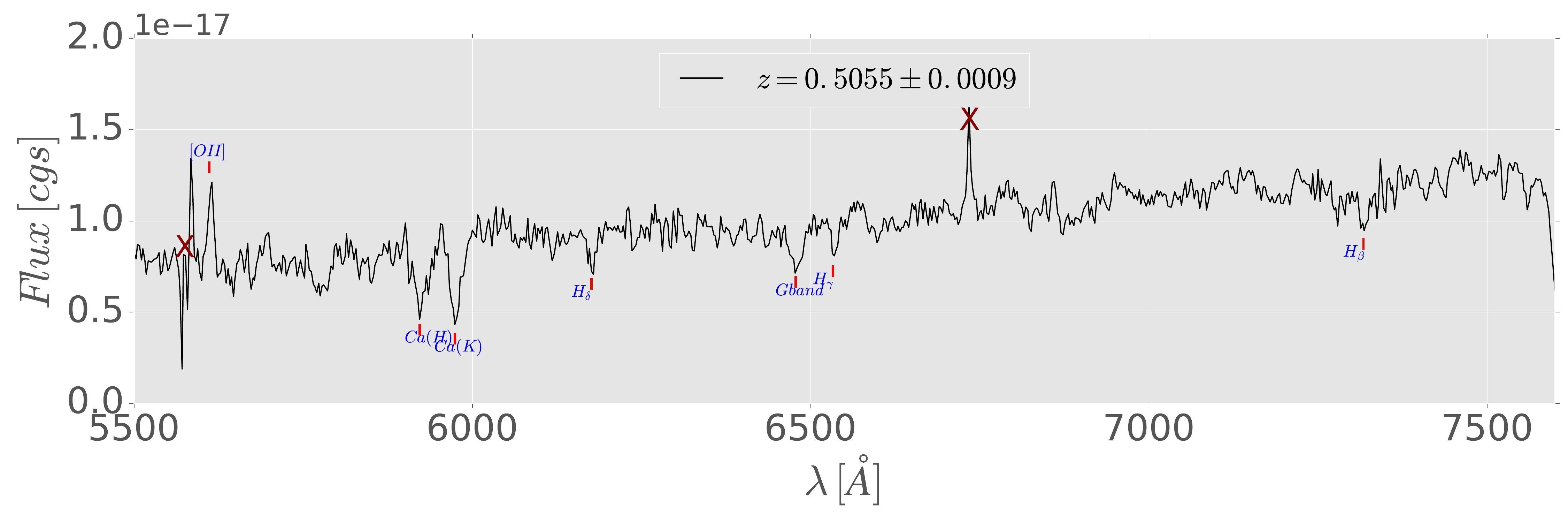}
\caption{Optical spectra of four of the 33 sources observed with GTC-OSIRIS. All spectra show emissions and/or absorption features, such as Balmer lines, oxygen  forbidden lines and calcium-II doublet at the observed wavelengths. The top-most panel shows the spectrum for the XID150 X-ray source at $z=0.6621$. Residual sky lines are marked by X sign.}
\label{spectra}
\end{figure*}

To reduce our spectroscopic data, we used the \texttt{gtcmos}\footnote{\url{https://www.inaoep.mx/~ydm/gtcmos/gtcmos.html}} package as explained in  \citet{Gomez2016},  a semi-automatic pipeline for the reduction of GTC/OSIRIS MOS data, which uses the standard IRAF tasks.  In Figure \ref{spectra}, we show an example of 4 spectra, which present typical signature of active galaxies (or star forming processes) such as  forbidden emission lines of  [\ion{O}{2}], and [\ion{O}{3}] (related to high level of ionization), intense permitted emission lines as  H$\alpha$, H$\beta$ and some absorption lines, such as CaK, CaH. Out of  $\sim$ 100 spectra, only 4 correspond to an X-ray counterpart, but only one of them, XID150 (top panel of Figure \ref{spectra}), presented clear  absorption/emission lines leading to a redshift estimate of  $z=0.6621$. All sources with spectroscopic redshift were included within the training set to compute the photo-{\it z} in section \ref{section3.3}.

\newpage
\section{Tables}
\label{AnexoA}

\begin{longrotatetable}
\begin{table*}
\tiny
\setlength\tabcolsep{3pt}
\centering
\caption{Main parameters of our catalog used along our X-ray to optical/IR analysis.}
\label{table5}
%\resizebox{\textwidth}{!}{%
\begin{tabular}{ccccccccccccccccccc}
\hline\hline
$IDX^a$ & $RA^b$ & $DEC^c$ & $ML^d$ & $Fx^e$ & $HR^f$ & $z^g$ & $zphot^h$ &  $Lx^i$ & $AngDis\_op^j$ & $g'^k$ & $r'^l$ & $i'^m$ & $W1^n$ & $W2^o$ & $Fr^p$ & $FWKs^q$ & $Log_{10}(Fx/Fr)^r$ & $log_{10}(Fx/Fks)^s$ \\\hline
1 & 238.9119 & 11.11235 & 104284.8 & 3.04E-13 & -0.5 & 2.6636 & 0.4 & 2.61E+46 & 1 & 19.42 & 19.31 & 19.27 & 15.97 & 15.23 & 4.99E-13 & 1.63E-13 & -0.214 & 0.271 \\
2 & 238.8756 & 11.36975 & 65311.24 & 1.44E-13 & -0.685 & 0.1343 & 0.15 & 7.14E+42 & 0.85 & 18.79 & 17.97 & 17.45 & 14.91 & 14.66 & 1.73E-12 & 4.30E-13 & -1.079 & -0.476 \\
3 & 238.7203 & 11.09557 & 39007.43 & 1.18E-13 & -0.51 & 0.756873 & 0.9 & 3.67E+44 & 0.55 & 19.39 & 19.54 & 19.37 & 14.73 & 13.49 & 4.04E-13 & 5.04E-13 & -0.534 & -0.631 \\
4 & 238.8523 & 11.15929 & 34690.83 & 8.55E-14 & -0.654 & -99 & 1.13 & 7.53E+44 & 0.6 & 20.67 & 20.43 & 20.38 & 16.2 & 15.23 & 1.79E-13 & 1.33E-13 & -0.32 & -0.19 \\
5 & 239.0276 & 11.27217 & 27177.42 & 8.41E-14 & -0.335 & 0.9481 & 0.58 & 4.70E+44 & 0.72 & 21.68 & 21.26 & 20.88 & -99 & -99 & 8.30E-14 & -99 & 0.006 & -99 \\
6 & 239.1048 & 11.16245 & 22552.52 & 6.42E-14 & -0.609 & -99 & 1.15 & 6.01E+44 & 0.71 & 21.26 & 20.91 & 20.66 & 16.92 & 15.73 & 1.15E-13 & 6.88E-14 & -0.252 & -0.031 \\
7 & 239.0026 & 11.06852 & 22232.23 & 7.08E-14 & -0.446 & -99 & 1.04 & 5.07E+44 & 0.91 & 21.51 & 21.32 & 21.12 & 15.95 & 15 & 7.88E-14 & 1.66E-13 & -0.047 & -0.369 \\
8 & 238.9871 & 11.29352 & 21306.9 & 5.50E-14 & -0.498 & -99 & 0.62 & 1.03E+44 & 0.53 & 22.45 & 22.23 & 21.51 & 17.66 & 16.82 & 3.40E-14 & 3.50E-14 & 0.208 & 0.196 \\
9 & 238.9754 & 10.99547 & 18809.03 & 7.78E-14 & -0.598 & -99 & 1 & 4.98E+44 & 0.36 & 20.7 & 20.51 & 20.37 & 15.86 & 15.72 & 1.66E-13 & 1.81E-13 & -0.328 & -0.366 \\
10 & 238.9107 & 11.34319 & 15904.15 & 6.18E-14 & -0.533 & 0.8788 & 0.79 & 2.84E+44 & 0.73 & 21.15 & 20.92 & 20.49 & 15.92 & 15.34 & 1.14E-13 & 1.70E-13 & -0.265 & -0.44\\

... & ... & ... & ... & ... & ... & ... & ... & ... & ... & ... & ... & ... & ... & ... & ... & ... & ... & ... \\ \hline \hline
\end{tabular}
%}
\begin{flushleft}
\tablecomments{\\ \footnotesize
$^a$ ID X-ray name for each source.
\\
$^b$ Right Ascension from the X-ray catalog.
\\
$^c$ Declination from the X-ray catalog.
\\
$^d$ Maximum Likelihood significance.
\\
$^e$ Full-band X-ray flux in units of $erg\,cm^{-2}\,s^{-1}$.
\\
$^f$ Hardness Ratio.
\\
$^g$ Spectroscopic redshift.  
\\
$^h$ Photometric redshift.
\\
$^i$ Full-band X-ray luminosity in units of $erg\, s^{-1}$.
\\
$^j$ Optical counterpart angular distant in  units of $arcsec$.
\\
$^k$ optical $g-band$  magnitude.
\\
$^l$ optical $r-band$ magnitude.
\\
$^m$ optical $i-band$ magnitude.
\\
$^n$ Infrared $W1-band$ magnitude.
\\
$^o$ Infrared $W2-band$ magnitude.
\\
$^p$ Optical flux at $r-band$ in units of $erg\, cm^{-2}\, s^{-1}$. 
\\
$^q$  Optical flux at $Ks-band$ in  units of $erg\, cm^{-2}\, s^{-1}$, expressed by the equation $Ks = 0.99\times W1 1+0.23$
\\
$^r$  Ratio between the X-ray and optical at r-band fluxes in $log_{10}$ scales 
\\
$^s$ Ratio between the X-ray and infrared at Ks-band fluxes in $log_{10}$ scales
\\
The whole table is available online in ASCII format along with this paper
}
\end{flushleft}
\end{table*}
\end{longrotatetable}

%\section{Figures}\label{AnexoB}
%\section{Author publication charges} \label{sec:pubcharge}

%% For this sample we use BibTeX plus aasjournals.bst to generate the
%% the bibliography. The sample63.bib file was populated from ADS. To
%% get the citations to show in the compiled file do the following:
%%
%% pdflatex sample63.tex
%% bibtext sample63
%% pdflatex sample63.tex
%% pdflatex sample63.tex

\bibliography{175UNDF}{}
\bibliographystyle{aasjournal}

%% This command is needed to show the entire author+affiliation list when
%% the collaboration and author truncation commands are used.  It has to
%% go at the end of the manuscript.
%\allauthors

%% Include this line if you are using the \added, \replaced, \deleted
%% commands to see a summary list of all changes at the end of the article.
%\listofchanges
%\turnoffeditone
%\turnoffedit
\end{document}